# Un índice discreto sensible a la desigualdad

Francisco José Zamudio Sánchez[1], Javier Jiménez Machorro*, Roxana Arana Ovalle**, Hildegardo Martínez Silverio***.

[1] Autor para correspondencia. fzamudios@chapingo.mx Profesor-Investigador en el Departamento de Estadística, Matemática y Cómputo de la Universidad Autónoma Chapingo. Texcoco, Estado de México, México. https://orcid.org/0000-0001-8252-9255
* Estudiante de la Maestría en Ciencias en Ciencias Forestales, en la Universidad Autónoma Chapingo.
** Universidad de Montreal.
*** Universidad Autónoma Chapingo.

## Abstract

This paper introduces the Relative Inequality Index at the Maximum (IDRM), a novel and intuitive measure designed to capture inequality within a population, such as income inequality. The index is based on the idea that individuals experience varying levels of inequality depending on their position within the distribution, particularly with respect to those at the top. The key assumption is that for individuals in lower positions, inequalities referenced to the top positions have greater impact on their well-being and the inequality relative to maximum is the most critical.

The IDRM fulfills desirable theoretical properties which were used for its evaluation and comparison against widely accepted measures in inequality literature. From this perspective, the IDRM is shown to be as robust as traditional measures and outperforms the Gini and Dalton indices by satisfying eight out of nine key properties, including decomposability across population subgroups. In a comparative analysis using income data from 58 countries and microdata from Mexico, with the Gini, Theil, and Atkinson indices as benchmarks, the IDRM demonstrates superior consistency, sensitivity to inequality, reduced bias in grouped data, and enhanced precision. This index reflects the varying forms of income distribution, showing heightened sensitivity to the magnitude of inequality.

## Resumen

Se propone un índice de desigualdad relativo al máximo (*IDRM*) como una nueva medida sencilla e intuitiva, que mide la desigualdad en los elementos de una población (ejemplo, distribución de ingresos). La idea elemental de su construcción es que cada individuo está expuesto a distintas desigualdades, con respecto a elementos en la parte inferior y superior en la distribución según su posición y que, sin duda, las que afectan su esfera de bienestar social son las segundas; de ahí que la desigualdad más importante es respecto a la posición más alta.

El índice propuesto cumple con propiedades teóricas deseables para las medidas de desigualdad, siendo éstas la base para su evaluación y contraste con los índices generalmente aceptados en la literatura sobre desigualdad. Desde esta perspectiva el índice muestra ser tan robusto como las medidas tradicionales y más robusto que el índice de Gini y Dalton, cumple ocho de las nueve propiedades, incluida la descomponibilidad en grupos de población.

En un análisis comparativo con datos de ingreso de 58 países y microdatos de México, tomando como referentes los índices Gini, Theil y Atkitson, el IDRM mostró características de congruencia, sensibilidad a la desigualdad, menor sesgo a datos agrupados y mayor precisión. El índice captura y expresa las distintas formas que puede adoptar la distribución del ingreso, siendo notoriamente sensible a la magnitud de las desigualdades, actuando como una herramienta sensible a la profundidad de las desigualdades y aproximando una idea del bienestar social.

**Statements and Declarations**

The authors did not receive support from any organization for the submitted work.

**Declaraciones y afirmaciones**

Los autores no recibieron apoyo de ninguna organización para el trabajo presentado.

## 1. Introducción

La desigualdad económica, sea por ingreso o por riqueza, tiene orígenes, probablemente, desde la aparición del hombre o, con mayor certeza, desde el inicio de la construcción de la cultura. Es tan antigua y crítica para las sociedades que innumerables trabajos existen sobre ella, sean teorías, factores que la afectan, consecuencias diversas que produce o conflictos sociales que desencadena. Lo mismo se podría decir de la desigualdad social, un atributo más general de la desigualdad.

La persistencia de esta condición de desigualdad, negativa para los países, es tal que pareciera ocioso pensar en su mejora. No obstante, hay lógicas formales e informales que indican la imposibilidad e inconveniencia económica de su eterna permanencia, al menos en los niveles que se han observado a través de la historia humana y su agudización en los últimos tiempos. Formalmente por los efectos de la desigualdad económica (en salud (Grant & O'Hara, 2010; Pickett & Wilkinson, 2015; Sapolsky, 2005), crecimiento económico (Alesina & Rodrik, 1994; Galor & Moav, 2004; Galor & Zeira, 1993), cohesión social (Leenders, R., 2014; Stigilitz, 2012), crimen (Corvalán & Pazzona, 2019; Daly et al., 2001; Kang, 2016; Kim et al., 2020), pobreza (Bernstein, 2014; Gould, 2014; Oxfam, 2013), vivienda (Matlack & Vigdor, 2008; Rodda, 1994; Vigdor, 2002), etcétera) no resulta conveniente para los Estados su permanencia. Informalmente se puede argumentar que los trabajadores y sus familias son los consumidores, de modo que si reciben poco, consumen poco y el crecimiento económico será cada vez menor, excepto que haya forma que unos cuantos, que poseen la mayor parte de la riqueza, puedan consumir todo lo que se produzca con los recursos que ellos mismos aportan para mantener el crecimiento.

La objetividad que se busca para tomar decisiones, o en actividades como la ciencia, hace necesario la medición de los atributos involucrados. Por ello, para el atributo desigualdad económica, implícito en varios procesos donde se deba decidir o se quiera confirmar una hipótesis sobre la desigualdad económica, se han desarrollado varias métricas (usualmente presentadas como índices), algunas se muestran en la primera sección. De ellas, tres son muy frecuentes en la literatura económica y demás ciencias sociales, los índices de Gini, Theil y Atkinson. Sin duda el índice de Gini es el más utilizado.

Cierto es que estos índices tienen especial interés en la desigualdad económica, pero se trasladan a cualquier tipo de atributo (esperanza de vida, ruralidad, salud, por nombrar algunos), donde la desigualdad tenga efectos adversos en la vida de las personas. En nuestro discurso (excepto en los ejemplos) nos referiremos a la desigualdad de ingresos en el sentido de Cowell (F. Cowell, 2011) pero es sólo por motivos de espacio.

La literatura sobre índices de desigualdad es amplia, un resumen referente es el proporcionado por Cowell (2011) página 151; la mayoría de los índices derivan o tienen una asociación con la curva de Lorenz, entre los más notables están el rango, la varianza, el coeficiente de variación al cuadrado, la varianza de los ingresos logarítmicos, las desviaciones medias absolutas y relativas, los índices de Atkinson, Theil y Gini (Heshmati, 2004).

En esta contribución se propone un índice (medida) de desigualdad sencillo e intuitivo, construido a partir de un razonamiento lógico que responde al sentido de desigualdad, entendido de que ésta sólo puede ser una y que las diferencias entre los valores del ingreso sólo representan grados de esa única desigualdad. En su fácil construcción se hicieron cambios que no se muestran por no ser relevantes, pero fueron hechos para buscar que el índice satisficiera el mayor número de propiedades que se han adoptado en la comunidad especializada en el tema, por ser deseable que una medida de desigualdad las cumpla.

La motivación del índice partió de apreciar que los índices actuales más usados, en general, son poco sensibles a los cambios que ocurren en el ingreso de las personas, más todavía entre aquellos en la parte superior y los de la parte más baja de la distribución. Hay coyunturas económicas, por ejemplo, ascensos muy significativos en las bolsas de valores donde no todos participan, de hecho lo hacen comúnmente los de mayores ingresos, y los índices más usados muestran cambios muy pequeños en la distribución de los ingresos. Se buscó, entonces, un índice que fuera sensible a los cambios que afectan la distribución de los ingresos. Aunque el índice no cambia cuando se hace una transferencia de un grupo con mayor ingreso a otro con menor ingreso, excepto cuando el primero es el grupo con mayor ingreso, resulta difícil concebir que haya grupos con mayor ventaja que estén dispuestos a transferir recursos a otros en desventaja.

La ventaja de este índice sobre los existentes, yace en su sensibilidad para expresar la brecha entre los que tienen los ingresos más altos y aquellos con los más bajos, además de la desigualdad que existe entre los demás ingresos con el mayor. Otra característica es la posibilidad de usarlo con grupos de ingresos de distintos tamaños poblacionales, no necesariamente por cuantiles. Adicionalmente, el índice captura movimientos en el ingreso del grupo más alto cuando los demás no sufren cambio o lo hacen en menor proporción (crece cuando la clase más alta incrementa su ingreso y los demás permanecen igual) y lo hace también cuando un determinado grupo incrementa su ingreso y los demás grupos permanecen igual (decrece cuando cualquier grupo, excepto el de mayor ingreso, incrementa su ingreso y los demás permanecen igual). Las dos últimas características se pueden considerar como propiedades adicionales que las medidas de desigualdad debieran poseer. Por último, el índice puede escribirse en función del número de Palma y explicar que tanto de la desigualdad deja de explicar este número.

El trabajo se estructura como sigue: sección 2) propiedades deseables de las medidas de desigualdad; 3) construcción del índice y propiedades que satisface; 4) resultados y discusión; 5) conclusiones.

## 2. Propiedades deseables de las medidas de desigualdad

En la literatura se han establecido diversas propiedades deseables para las medidas de desigualdad, las cuales sirven para evaluarlas o compararlas. Enseguida se presenta una lista de las principales (Atuesta et al., 2018; F. A. Cowell & Victoria-Feser, 1996; Mader, 2000):

• P1. Anonimidad o simetría: Si cualquier par de unidades en la población intercambian su nivel de ingreso, el índice de desigualdad no varía.

Esta propiedad implica que el índice se calcula exclusivamente sobre la base del vector de observaciones del ingreso, sin que sean relevantes otras características de los individuos.

•P2. Invarianza a la escala (homogeneidad de grado cero, o independencia de media): Si la variable de análisis se multiplica por el mismo escalar para todos los individuos de la población, el grado de desigualdad no varía.

Esta propiedad, implícita en la mayoría de los índices de desigualdad, implica que la desigualdad se cuantifique de manera relativa, es decir, tomando como referencia el nivel promedio (o un umbral) de la variable de interés. Esta propiedad permite además que el grado de desigualdad no dependa de la unidad de medida en que se expresa una variable (por tanto, no es relevante si los ingresos, por ejemplo, se expresan en pesos, miles de pesos o dólares).

•P3. Invarianza a las réplicas (independencia de población): Si la población se replica o repite un número finito de veces, el índice de desigualdad no varía.

Esta propiedad permite que los resultados del índice sean comparables entre poblaciones de distinto tamaño.

• P4. Principio de transferencias (o condición de Dalton-Pigou): Transferencias de ingreso de unidades de la parte alta de la distribución a la parte baja de la distribución mantienen (condición débil) o reducen (condición fuerte) la medida de desigualdad.

El principio de Dalton-Pigou resume la característica principal de un índice de desigualdad, que lo diferencia de la mayoría de índices estadísticos de dispersión. Esta propiedad conlleva a que un índice de desigualdad debe asignar ponderaciones distintas a los ingresos, según el lugar en el que se encuentren en la distribución de los mismos.

• P5. Principio de sensibilidad a transferencias: Si se tienen dos pares de unidades, uno relativamente más rico y el otro más pobre, separados por la misma distancia de ingresos, una transferencia progresiva reducirá la desigualdad más en el segundo par que en el primero (condición fuerte), o al menos la desigualdad será la misma en ambos pares (condición débil)[2].

• P6. No negatividad: la métrica de desigualdad debe ser mayor o igual a cero.

• P7. Cero igualitario: La métrica toma el valor cero en el caso igualitario, cuando todas las unidades en la población tienen el mismo ingreso.

• P8. Acotado superiormente por máxima desigualdad: La métrica de desigualdad alcanza su valor máximo, de máxima desigualdad cuando todas las unidades tienen el valor cero, excepto una. Este valor suele ser la unidad cuando el número de unidades $n$ se acerca a infinito.

• P9. Descomponible por subgrupos: La medida de desigualdad se puede descomponer por subgrupos de población. Es particularmente deseable que la separabilidad sea aditiva, es decir, que el valor del índice para toda la población pueda obtenerse como la suma de las desigualdades intra-grupales e inter-grupales de los subgrupos utilizados. Una referencia obligada en el estudio de la descomposición por subgrupos, es el trabajo de Bourguignon (Bourguignon, 1979).

---

[2] El principio de sensibilidad a transferencias puede ser ilustrado mediante el siguiente ejemplo numérico. Supóngase cuatro unidades (denotadas A, B, C y D), con ingresos de $10, $20, $30 y $40. Supóngase una transferencia de $2 de la unidad B a la unidad A. La nueva distribución será $12, $18, $30, $40. Un índice que cumpla con el principio de Dalton-Pigou dará como resultado de esta transferencia una medida de desigualdad igual(débil) o menor(fuerte). Ahora bien, volviendo a la distribución original, supóngase una transferencia de $2 de la unidad D a la unidad C. La distribución resultante será $10, $20, $32, $38. Conforme al principio de Dalton-Pigou, esta distribución también tendrá una medida de desigualdad menor o igual que la primera. El principio de sensibilidad a transferencias requiere, adicionalmente al principio de Dalton-Pigou, que la segunda distribución tenga una menor medida de desigualdad que la tercera (condición fuerte), o al menos tendrán iguales medidas de desigualdad (condición débil).

## 3. Construcción del índice y propiedades que satisface

### 3.1. Definición

Notación:

- La población objeto de estudio está representada por $n$ unidades (individuos, municipios, regiones, países, etc.) Cada unidad *i* contiene $m_i$ elementos, *i* = 1, 2, …, *n*, y denotamos por $m_\bullet = \sum_1^n m_i$. Es decir, cada unidad puede tener uno o más elementos.
- El ingreso (característica o atributo de interés) de cada elemento en la unidad $i$ es $x_i$, $i = 1, \dots, n$. Considere $\underline{x} = (x_1, x_2, \dots, x_n)'$.
- La proporción poblacional en la unidad $i$ es $p_i = {}^{m_i}/_{m_\bullet}$, $i = 1, \dots, n$. Sea $\underline{p} = (p_1, p_2, \dots, p_n)'$.
- El ingreso máximo observado es $x_{máx} = máx\{x_1, x_2, \dots, x_n\} = máx(\underline{x})$.

El valor de desigualdad asociado a la unidad $i$ es el diferencial entre el valor máximo observado y el valor de la unidad $i$, relativo al valor máximo:

$$d_i = \frac{x_{máx} - x_i}{x_{máx}}$$

Se considera esta diferencia en la unidad *i* por ser única y la de mayor interés. Para aclarar lo anterior, supongamos los valores de las unidades *u* y *v*, tales que:

$$x_v < x_i < x_u < x_{máx}$$

Las otras diferencias positivas que involucran a la unidad *i* son $(x_u - x_i)$ y $(x_i - x_v)$, la primera es menor a la que se tiene con el máximo y de considerarse esa fracción se duplicaría con la que ya está considerada respecto al máximo; situación análoga resulta con la segunda, ya que, es menor a la que se considerará para el valor de la unidad *v*, $(x_{máx} - x_v)$, y la fracción $(x_i - x_v)$ se duplicaría con la anterior. Así, el valor de desigualdad asociado a la unidad *i*, $(x_{máx} - x_i)$, es el único relevante para ella, respecto a los que son mayores, $(x_u - x_i)$, ya que, de no existir la primera diferencia, no existirían tampoco las segundas; mientras que las diferencias con valores menores, $(x_i - x_v)$, no existirían si la desigualdad $(x_{máx} - x_v)$ fuera nula.

En este trabajo se propone que el valor de la desigualdad en la población objeto de estudio, $(nt)_{\underline{x}}$, sea el resultado de agregar el valor de la desigualdad de todas las unidades, ponderadas por la proporción poblacional que representan:

$$(nt)_{\underline{x}} = \sum_{i=1}^{n} p_i d_i = \sum_{i=1}^{n} p_i \left(\frac{x_{máx}-x_i}{x_{máx}}\right) = \sum_{i=1}^{n} p_i \left(1 - \frac{x_i}{x_{máx}}\right) = \sum_{i=1}^{n} p_i - \sum_{i=1}^{n} \frac{p_i x_i}{x_{máx}}$$

$$= \underline{p}'\underline{1}_n - \frac{\underline{p}'\underline{x}}{x_{máx}} = \underline{p}'\left(\underline{1}_n - \frac{\underline{x}}{x_{máx}}\right) = \underline{p}'\left(\underline{1}_n - \frac{\underline{x}}{máx(\underline{x})}\right) = 1 - \frac{\underline{p}'\underline{x}}{máx(\underline{x})}$$

En adelante $(nt)_{\underline{x}}$ será referido como Índice de Desigualdad Relativa al Máximo (*IDRM*) y observe que su valor es calculado de la matriz $[\underline{x}\ \underline{p}]$[3].

### 3.2. Propiedades

El *IDRM* cumple con 8 propiedades deseables de las medidas de desigualdad, y no cumple la condición fuerte del Principio de Sensibilidad a Transferencias en ningún caso.

#### 3.2.1. Simetría o anonimidad

Si cualquier par de individuos en distintas unidades intercambian sus ingresos, la desigualdad no cambia. Si un individuo de la unidad i intercambia su ingreso con otro de la unidad j, es claro que la matriz $\left[\underline{x}\ \underline{p}\right]$ no cambia, por tanto $(nt)_{\underline{x}}$ permanece igual.

#### 3.2.2. Invarianza a la escala u homogeneidad de grado cero

Sea $\alpha > 0$ tal que $\alpha\underline{x}$ representa un cambio de escala sobre $\underline{x}$, es decir, que el atributo y proporciones de población en estudio están dadas por $\left[\underline{x}\ \underline{p}\right]\begin{bmatrix}\alpha & 0\\ 0 & 1\end{bmatrix} = [\alpha\underline{x}\ \underline{p}]$, entonces el valor de la medida de desigualdad $(nt)_{\underline{x}}$ no cambia, ya que:

$$(nt)_{\alpha\underline{x}} = \underline{p}'\left(\underline{1}_n - \frac{\alpha\,\underline{x}}{máx(\alpha\,\underline{x})}\right) = \underline{p}'\left(\underline{1}_n - \frac{\alpha\,\underline{x}}{\alpha\,máx(\underline{x})}\right) = (nt)_{\underline{x}}$$

#### 3.2.3. Invarianza a las réplicas

Si se une la población en $\underline{x}$ con una copia de ella misma, $\underline{x} \cup \underline{x}$ , ahora cada $x_i$ estaría ligada a $2m_i$ elementos, donde $m_i$ son los elementos en la unidad $i$, así el total de elementos cambia de $m_\bullet = \sum_{i=1}^{n} m_i$ a $2m_\bullet = 2\sum_{i=1}^{n} m_i$ y la nueva $p_i = \frac{2m_i}{2m_\bullet} = \frac{m_i}{m_\bullet}$, lo cual no modifica el valor de $(nt)_{\underline{x}}$. Note que los valores en $\underline{x}$ no cambian.

[3] En la etapa final de redacción de este manuscrito, se identificó que la forma del ***IDRM***, que postulamos como medida de **desigualdad** y que construimos analizando cuál es la desigualdad más severa a la que están sometidos todos los individuos en una población, puede verse como una adaptación de un caso particular de la familia de índices para medir **pobreza,** propuesta por Foster, J., Greer, J., & Thorbecke, E. (1984).

### 3.2.4. Principio de transferencias o condición Dalton – Pigou

***Condición débil***

Sea $\underline{x}_{()}$ el vector que contiene las componentes ordenadas de menor a mayor del vector $\underline{x}$, denotadas por $x_{(i)}$. A las correspondientes proporciones $p_i$, las denotaremos por $p_{(i)} = {}^{m_{(i)}}/_{m_{\bullet}}$ y al vector que las contiene $\underline{p}_{()}$. La matriz $[\underline{x}_{()}\ \underline{p}_{()}]$ es un arreglo (permutaciones) de las hileras de la matriz $[\underline{x}\ \underline{p}]$. En esta nueva notación debe observarse que $x_{máx} = x_{(n)}$ y $(nt)_{\underline{x}} = (nt)_{\underline{x}_{()}}$.

Considere una transferencia de tamaño $\varepsilon$ (infinitesimal) de cada uno de los elementos de la unidad $(j)$ (transferencia total = $m_{(j)}\varepsilon$) en partes iguales a cada uno de los elementos de la unidad $(i)$ (cada elemento en $(i)$ recibe = $\frac{m_{(j)}}{m_{(i)}}\varepsilon = \frac{p_{(j)}}{p_{(i)}}\varepsilon$), donde $x_{(i)} < x_{(j)} \neq x_{(n)}$ y la transferencia no modifica el orden de $\underline{x}_{()}$.

Después de la transferencia, las coordenadas de $\underline{x}_{()}$ serán las mismas, excepto la *i*-ésima y la *j*-ésima; la *i*-ésima cambiará a $x_{(i)} + \frac{p_{(j)}}{p_{(i)}}\varepsilon$, y la *j*-ésima a $x_{(j)} - \varepsilon$. Denotemos por $\underline{0}_{n_{-(i),(j)}}(k, q)$ a un vector de orden *n* donde todas sus coordenadas son cero, excepto la (*i*)-ésima con valor *k*, y la (*j*)-ésima con valor *q*. El nuevo vector de ingresos se puede escribir como $\underline{y}_{()} = \underline{x}_{()} + \underline{0}_{n_{-(i),(j)}}(\frac{p_{(j)}}{p_{(i)}}\varepsilon, -\varepsilon)$, el cual tiene un *IDRM*:

$$(nt)_{\underline{y}_{()}} = \underline{p}_{()}'\left(\underline{1}_n - \frac{\underline{y}_{()}}{máx(\underline{y}_{()})}\right) = \underline{p}_{()}'\left(\underline{1}_n - \frac{\underline{x}_{()} + \underline{0}_{n_{-(i),(j)}}(\frac{p_{(j)}}{p_{(i)}}\varepsilon, -\varepsilon)}{máx(\underline{x}_{()})}\right)$$

$$= \underline{p}_{()}'\left(\underline{1}_n - \frac{\underline{x}_{()}}{máx(\underline{x}_{()})}\right) = (nt)_{\underline{x}_{()}} = (nt)_{\underline{x}}$$

En la segunda igualdad se usó $máx\left(\underline{y}_{()}\right) = máx(\underline{x}_{()})$, porque la transferencia no cambia el orden de los ingresos y en la tercera igualdad se ocupó $\underline{p}_{()}'\underline{0}_{n_{-(i),(j)}}\left(\frac{p_{(j)}}{p_{(i)}}\varepsilon, -\varepsilon\right) = 0$, ya que todos los términos del producto escalar son cero, excepto el (*i*)-ésimo, $p_{(i)}\frac{p_{(j)}}{p_{(i)}}\varepsilon$, y el *j*-ésimo, $-p_{(j)}\varepsilon$, los cuales se cancelan.

Las anteriores ecuaciones indican que el *IDRM* es invariante a transferencias de una unidad con ingreso más alto a una unidad con ingreso más bajo, de modo que se cumple la condición débil de Dalton – Pigou.

***Condición fuerte***

Para el caso donde la unidad $(j)$ de la cual se realizará la transferencia, sea $(n)$, la de mayor ingreso, la condición fuerte de Dalton – Pigou se cumple, ya que:

$$(nt)_{\underline{y}_{()}} = \underline{p}_{()}'\left(\underline{1}_n - \frac{\underline{y}_{()}}{máx(\underline{y}_{()})}\right) = \underline{p}_{()}'\left(\underline{1}_n - \frac{\underline{x}_{()} + \underline{0}_{n-(i),(n)}(\frac{p_{(n)}}{p_{(i)}}\varepsilon, -\varepsilon)}{x_{(n)} - \varepsilon}\right)$$

$$= \underline{p}_{()}'\left(\underline{1}_n - \frac{\underline{x}_{()}}{x_{(n)} - \varepsilon}\right)$$

$$= \underline{p}_{()}'\left(\underline{1}_n - \frac{\underline{x}_{()}}{x_{(n)} - \varepsilon} - \frac{\underline{x}_{()}}{x_{(n)}} + \frac{\underline{x}_{()}}{x_{(n)}}\right) = (nt)_{\underline{x}_{()}} - \underline{p}_{()}'\underline{x}_{()}\left(\frac{1}{x_{(n)} - \varepsilon} - \frac{1}{x_{(n)}}\right)$$

$$= (nt)_{\underline{x}} - \underline{p}_{()}'\underline{x}_{()}\left(\frac{\varepsilon}{x_{(n)}(x_{(n)} - \varepsilon)}\right) < (nt)_{\underline{x}}$$

### 3.2.5. Principio de sensibilidad a transferencias

Como se observó en 3.2.4., una transferencia de un ingreso alto a otro bajo, no modifica el *IDRM*, no importa si el par de unidades involucradas están en la parte alta de la distribución de ingresos o en la parte baja. Consecuentemente, el *IDRM* no cumple el principio fuerte de sensibilidad a transferencias pero sí el principio débil.

### 3.2.6. No negatividad:

$$(nt)_{\underline{x}} = \sum_{i=1}^{n} p_i\left(\frac{x_{máx} - x_i}{x_{máx}}\right) \geq 0$$

### 3.2.7. Cero igualitario

Note que para $k > 0$ y $\underline{x} = k\underline{1}_n$ entonces $(nt)_{\underline{x}=k\underline{1}_n} = \underline{p}'\left(\underline{1}_n - \frac{\underline{x}}{máx(\underline{x})}\right) = \underline{p}'\left(\underline{1}_n - \frac{k\underline{1}_n}{k}\right) = 0$

### 3.2.8. Acotado por arriba por máxima desigualdad

Si $\forall\, i$, excepto $i^*$, $x_i = 0$ y $x_{i^*} = k$, con $k > 0$, entonces si denotamos por $\underline{0}_{n_{-i}}(k)$ al vector de orden $n$ con ceros en todas sus coordenadas excepto la i-ésima con valor igual a $k$, se tiene que, en este caso, $\underline{x} = \underline{0}_{n_{-i^*}}(k)$ y,

$$(nt)_{\underline{x}} = \underline{p}'\left(\underline{1}_n - \frac{\underline{x}}{\text{máx}(\underline{x})}\right) = \underline{p}'\left(\underline{1}_n - \frac{\underline{0}_{n_{-i^*}}(k)}{\text{máx}(\underline{x})}\right) = \underline{p}'\underline{1}_n - \frac{kp_{i^*}}{k} = 1 - p_{i^*}$$

$$= 1 - \frac{m_{i^*}}{m_{\bullet}} \longrightarrow 1\; cuando\; n \longrightarrow \infty, ya\; que, m_{\bullet} \longrightarrow \infty$$

$$(nt)_{\underline{x}} = 1 - \frac{p_{i^*}k}{k} = 1 - p_{i^*} = 1 - \frac{n_{i*}}{n} \longrightarrow 1\; cuando\; n \longrightarrow \infty.$$

### 3.2.9. Descomponible por grupos

Es deseable que una medida de desigualdad se pueda descomponer por subgrupos de población. Particularmente que la separabilidad sea aditiva, es decir, que el valor del índice para toda la población pueda obtenerse como la suma de las desigualdades intra-grupales e inter-grupales de los subgrupos utilizados. El *IDRM* puede descomponerse de manera aditiva.

Suponga que la composición de $\left[\underline{x}\ \underline{p}\right]$ proviene de una partición de la población con $g$ grupos y cada grupo $k$ tiene $n_k$ unidades, $k = 1, 2, \dots, g$, de modo que $\sum_{k=1}^{g} n_k = n$.

El ingreso en la unidad $i$ que está en el grupo $k$ se denota por $x_i(k)$ y la proporción de la población que representa este valor por $p_i(k)$, $i = 1, \dots, n_k$.

El valor máximo observado del atributo en el grupo $k$ se denota por $x_{máx}{}^{(n_k)}$ y el valor máximo global por $x_{máx}{}^{(n)}$.

La desigualdad asociada al individuo $i$ del grupo $k$ es:

$$n_i(k) = \frac{x_{máx}{}^{(n_k)} - x_i(k)}{x_{máx}{}^{(n_k)}}$$

Por lo tanto, el *IDRM* del grupo $k$, denotado aquí por $IDRM(k)$ es:

$$IDRM(k) = \sum_{i=1}^{n_k} p_i(k) n_i(k) = \sum_{i=1}^{n_k} p_i(k)\left[1 - \frac{x_i(k)}{x_{máx}{}^{(n_k)}}\right] = \sum_{i=1}^{n_k} p_i(k) - \frac{\sum_{i=1}^{n_k} p_i(k) x_i(k)}{x_{máx}{}^{(n_k)}}$$

$$= s(k) - \frac{x_{máx}{}^{(n)}}{x_{máx}{}^{(n_k)}} \sum_{i=1}^{n_k} \frac{p_i(k)x_i(k)}{x_{máx}{}^{(n)}},$$ donde $s(k) = \sum_{i=1}^{n_k} p_i(k)$.

De modo equivalente,

$$\frac{x_{máx}{}^{(n)}}{x_{máx}{}^{(n_k)}} \sum_{i=1}^{n_k} \frac{p_i(k)x_i(k)}{x_{máx}{}^{(n)}} = s(k) - IDRM(k) \Leftrightarrow$$

$$\sum_{i=1}^{n_k} \frac{p_i(k)x_i(k)}{x_{máx}{}^{(n)}} = \frac{x_{máx}{}^{(n_k)}}{x_{máx}{}^{(n)}}\, s(k) - \frac{x_{máx}{}^{(n_k)}}{x_{máx}{}^{(n)}}\, IDRM(k) \Leftrightarrow$$

$$-\sum_{i=1}^{n_k} \frac{p_i(k)x_i(k)}{x_{máx}{}^{(n)}} = \frac{x_{máx}{}^{(n_k)}}{x_{máx}{}^{(n)}}\, IDRM(k) - \frac{x_{máx}{}^{(n_k)}}{x_{máx}{}^{(n)}} s(k)$$

Entonces, la desigualdad total puede escribirse como:

$$(nt)_{\underline{x}} = 1 - \sum_{k=1}^{g} \sum_{i=1}^{n_k} \frac{p_i(k)x_i(k)}{x_{máx}{}^{(n)}} = \sum_{k=1}^{g} \frac{x_{máx}{}^{(n_k)}}{x_{máx}{}^{(n)}}\, IDRM(k) - \sum_{k=1}^{g} \frac{x_{máx}{}^{(n_k)}}{x_{máx}{}^{(n)}} s(k) + 1$$

$$= \sum_{k=1}^{g} \frac{x_{máx}{}^{(n_k)}}{x_{máx}{}^{(n)}}\, IDRM(k) + \left[1 - \sum_{k=1}^{g} \frac{x_{máx}{}^{(n_k)}}{x_{máx}{}^{(n)}} s(k)\right] \qquad (1)$$

Por otro lado, $\sum_{k=1}^{g} s(k) = \sum_{k=1}^{g} \sum_{i=1}^{n_k} p_i(k) = 1$, ya que es la suma de las proporciones de todas las unidades en la población. Por lo tanto

$$\left[1 - \sum_{k=1}^{g} \frac{x_{máx}{}^{(n_k)}}{x_{máx}{}^{(n)}}\, s(k)\right] = \sum_{k=1}^{g} s(k) - \sum_{k=1}^{g} \frac{x_{máx}{}^{(n_k)}}{x_{máx}{}^{(n)}}\, s(k) = \sum_{k=1}^{g} s(k) \left[1 - \sum_{k=1}^{g} \frac{x_{máx}{}^{(n_k)}}{x_{máx}{}^{(n)}}\right] = (nt)_{\underline{x}_{máx}}, \qquad (2)$$

donde $\underline{x}_{máx} = (x_{máx}{}^{(n_1)}, x_{máx}{}^{(n_2)}, \ldots, x_{máx}{}^{(n_g)})$. Es decir, $(nt)_{\underline{x}_{máx}}$ es el *IDRM* entre grupos, el cual podemos renombrar como $IDRM(\underline{x}_{máx})$, es decir, $(nt)_{\underline{x}_{máx}} = IDRM(\underline{x}_{máx})$.

Substituyendo la ecuación (2) en la (1), obtenemos la descomposición por grupos:

$$(nt)_{\underline{x}} = \sum_{k=1}^{g} \frac{x_{máx}{}^{(n_k)}}{x_{máx}{}^{(n)}}\, IDRM(k) + \left[1 - \sum_{k=1}^{g} \frac{x_{máx}{}^{(n_k)}}{x_{máx}{}^{(n)}} s(k)\right]$$
$$= \sum_{k=1}^{g} \frac{x_{máx}{}^{(n_k)}}{x_{máx}{}^{(n)}}\, IDRM(k) + IDRM(\underline{x}_{máx})$$

Así, el *IDRM* total, $(nt)_{\underline{x}}$, es una combinación lineal de los *IDRM* dentro de cada grupo, $\sum_{k=1}^{g} \frac{x_{\text{máx}}^{(n_k)}}{x_{\text{máx}}^{(n)}} \, IDRM(k)$, más el *IDRM* entre grupos, $IDRM(\underline{x}_{\text{máx}})$.

### 3.2.10. Otras propiedades

#### *Una función de bienestar social asociada al IDRM*

Si $x$ es el ingreso de un individuo en una distribución donde el máximo ingreso se representa por $x_{\text{máx}}$, entonces podemos definir una función de utilidad individual $U(x) = x/x_{\text{máx}}$. Observe que tal utilidad está definida en términos de ingresos pero no es un ingreso, puede interpretarse como la satisfacción del consumidor con ingreso $x$ respecto a la que obtiene el consumidor con ingreso $x_{\text{máx}}$ en la obtención de productos, lo cual atiende una crítica a las funciones de utilidad expresadas como ingresos y no como medidas de satisfacción (Magdalou, 2018).

Con la anterior función de utilidad individual, la de bienestar social estaría dada por:

$$U(\underline{x}_{()}) = \frac{1}{m_\bullet} \sum_{i=1}^{n} \frac{m_{(i)} x_{(i)}}{x_{(n)}} = \sum_{i=1}^{n} \frac{p_{(i)} x_{(i)}}{x_{(n)}} = \frac{\bar{x}}{x_{(n)}}$$

Esta función de bienestar está acotada, toma valores desde $p_n$ (todas las unidades, excepto una, tienen ingreso cero) hasta 1 (todas las unidades reciben el mismo ingreso medio $\sum_{i=1}^{n} \frac{m_{(i)} x_{(i)}}{m_\bullet} = \sum_{i=1}^{n} p_{(i)} x_{(i)}$).

#### *La curva de Lorenz y el IDRM*

El *IDRM* es función del área bajo la curva de Lorenz como se puede observar de la siguiente representación del *IDRM*. Denotemos por $T = \sum_{i=1}^{n} m_{(i)} x_{(i)}$ al ingreso total, entonces:

$$(nt)_{\underline{x}} = (nt)_{\underline{x}_{()}} = \sum_{i=1}^{n} p_{(i)} \left( \frac{x_{(n)} - x_{(i)}}{x_{(n)}} \right) = 1 - \frac{\sum_{i=1}^{n} p_{(i)} \, x_{(i)}}{x_{(n)}} = 1 - \frac{1}{m_\bullet x_{(n)}} \sum_{i=1}^{n} m_{(i)} x_{(i)}$$

$$= 1 - \frac{T}{m_\bullet x_{(n)}} \sum_{i=1}^{n} \frac{m_{(i)} x_{(i)}}{T} = 1 - \frac{T}{m_\bullet x_{(n)}} \sum_{i=1}^{n} s_{(i)} = 1 - \sum_{i=1}^{n} \delta s_{(i)}$$

donde $\delta = \frac{T}{m_\bullet x_{(n)}}$ y $s_{(i)}$ es la cuota de ingreso correspondiente a la unidad $(i)$. Si el *j*-ésimo punto de la curva de Lorenz es $(\sum_{i=1}^{i=j} p_{(i)}, \sum_{i=1}^{i=j} s_{(i)})$, $j = 1, 2, \dots, n$, entonces, el *j*-ésimo punto de la curva vinculada al *IDRM* es $(\sum_{i=1}^{i=j} p_{(i)}, \sum_{i=1}^{i=j} \delta s_{(i)})$, $j = 1, 2, \dots, n$, (Gráfica 1). En otras palabras, el *IDRM* es el área del cuadrado unitario menos el área de la curva de Lorenz vinculada al *IDRM*.

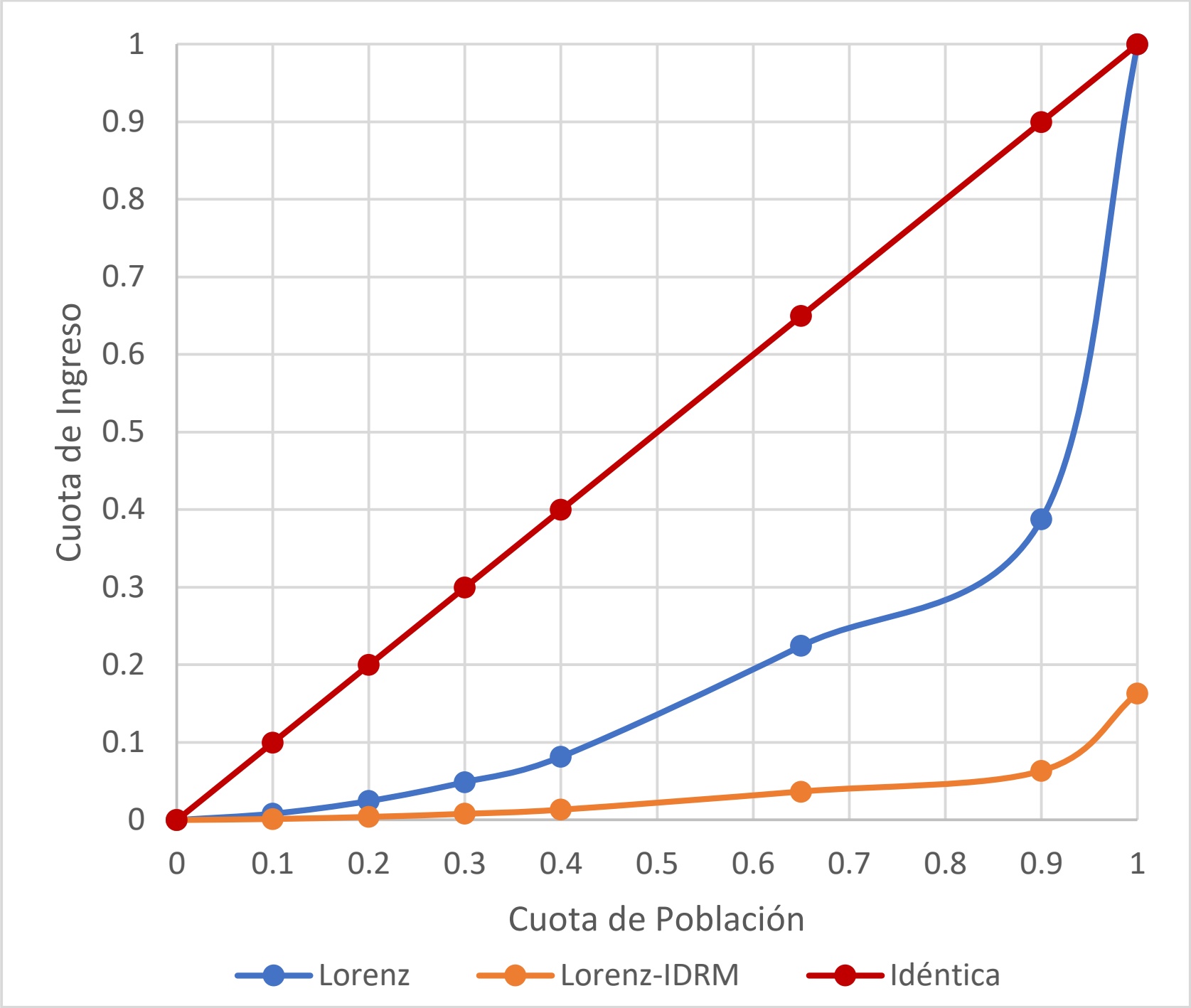

*Gráfica 1. Curva de Lorenz e IDRM*

**Fuente: Elaboración propia.**

***El índice de Atkinson ($I_A$) y el IDRM***

El índice de Atkinson se puede expresar como:

$$I_A\left(\underline{x}, \varepsilon\right) = 1 - \frac{x_{EDE}}{\bar{x}}$$

Donde $x_{EDE} = \left[\frac{1}{m_\bullet}\sum_{i=1}^{n} m_{(i)} x_{(i)}^{1-\varepsilon}\right]^{\frac{1}{1-\varepsilon}}$, es el ingreso igualmente distribuido equivalente con parámetro de aversión a la desigualdad igual a $\varepsilon$, y $\bar{x} = \sum_{i=1}^{n} p_{(i)} x_{(i)}$ es el ingreso medio (Liberati & Bellù, 2006).

Asimismo, el *IDRM* puede escribirse como:

$$(nt)_{\underline{x}} = (nt)_{\underline{x}_{()}} = \sum_{i=1}^{n} p_{(i)} \left(\frac{x_{(n)} - x_{(i)}}{x_{(n)}}\right) = 1 - \frac{\sum_{i=1}^{n} p_{(i)}\, x_{(i)}}{x_{(n)}} = 1 - \frac{\bar{x}}{x_{(n)}}$$

### *Definición de un parámetro de tolerancia a la desigualdad ($\tau$) para el IDRM*

$x_{(n)}$ puede escribirse en función de $\bar{x}$:

$$\bar{x}=\sum_{i=1}^{n}p_{(i)}\,x_{(i)}=\sum_{i=1}^{n}p_{(i)}\,f_{(i)}x_{(n)}\Leftrightarrow x_{(n)}=\frac{\bar{x}}{\sum_{i=1}^{n}p_{(i)}\,f_{(i)}}$$

Donde $f_{(i)}=\frac{x_{(i)}}{x_{(n)}}, i=1,2,\ldots,n$. Es claro que $1/\sum_{i=1}^{n}p_{(i)}\,f_{(i)}\geq 1$, si fuera igual a 1, no habría desigualdad. Se puede definir como parámetro de tolerancia a la desigualdad, denotada por $\tau$, a la diferencia:

$$\tau=\frac{1}{\sum_{i=1}^{n}p_{(i)}\,f_{(i)}}-1=\frac{x_{(n)}}{\frac{\sum_{i=1}^{n}m_{(i)}\,x_{(i)}}{m_{\bullet}}}-1=\frac{x_{(n)}}{\bar{x}}-1$$

Los valores de $\tau\in[0,\infty)$ y vale $t$ cuando $x_{(n)}=(t+1)\bar{x}$.

### *Definición de un ingreso más igualmente distribuido equivalente ($x_{MIDE}$) para el IDRM*

Dada la función de bienestar social definida para el *IDRM*, no existe un ingreso igualmente distribuido equivalente, ya que, para mantener la misma función de bienestar el ingreso $x_{(n)}$ debe permanecer; no obstante, se puede definir un ingreso más igualmente distribuido equivalente, denotado por $x_{MIDE}$, como sigue. Considere el mismo ingreso $x_{MIDE}$ para todos, excepto uno de la unidad con el máximo ingreso $x_{(n)}$ ($m_{\bullet}-1$ de los elementos en la población reciben $x_{MIDE}$ y 1 de ellos recibe $x_{(n)}$), de modo que esta distribución tenga el mismo bienestar social que la original, es decir, se cumpla:

$$\frac{1}{m_{\bullet}}\frac{(m_{\bullet}-1)x_{MIDE}+\ x_{(n)}}{x_{(n)}}=\frac{1}{m_{\bullet}}\sum_{i=1}^{n}\frac{m_{(i)}x_{(i)}}{x_{(n)}}=U\big(\underline{x}_{()}\big)=\frac{1}{m_{\bullet}x_{(n)}}\left[\sum_{i=1}^{n}m_{(i)}x_{(i)}\right]=\frac{1}{m_{\bullet}x_{(n)}}T$$

Lo anterior es equivalente a:

$$(m_{\bullet}-1)x_{MIDE}+\ x_{(n)}=T=m_{\bullet}\bar{x}$$

Así:

$$x_{MIDE}=\frac{m_{\bullet}\bar{x}-x_{(n)}}{m_{\bullet}-1}$$

Este ingreso llama la atención por ser casi $\bar{x}$ cuando la población es grande. Imagine una población con 126'014,024 elementos ($m_{\bullet}$), cuyo ingreso medio sea de 50,309.00 ($\bar{x}$) pero una persona tenga un ingreso de 10'702,107 ($x_{(n)}$), entonces $x_{MIDE} = 50{,}308.916$. Para esta situación el $IDRM = (nt)_{\underline{x}} = (nt)_{\underline{x}_{()}} = 1 - \frac{\bar{x}}{x_{(n)}} = 0.9953$, la tolerancia a la desigualdad $\tau = \frac{x_{(n)}}{\bar{x}} - 1 = 211.73$ y el valor de la función de bienestar social $U\left(\underline{x}_{()}\right) = \frac{\bar{x}}{x_{(n)}} = 0.0047$. Es decir, si 126'014,023 elementos recibieran el ingreso más igualmente distribuido equivalente de 50,308.916 y un elemento recibiera 10'702,107, habría una desigualdad casi perfecta, una función de bienestar casi de cero con una tolerancia a la desigualdad muy grande. Por otro lado, si todos en la población recibieran 50,309.00, no habría desigualdad, la tolerancia sería cero y el bienestar tendría el valor más alto posible (1). Así, una diferencia de aproximadamente 0.084 en el ingreso de cada elemento, la cual retiene una sola persona, induce un gran desequilibrio en la igualdad (¡y el mercado!), 0.9953, y un bienestar muy bajo, 0.0047, a cambio de una tolerancia muy alta, 211.73. Los datos anteriores corresponden a la Encuesta Nacional de Ingresos y Gastos en los Hogares de México para el ingreso trimestral en 2020 (*Encuesta Nacional de Ingresos y Gastos de los Hogares (ENIGH). 2020 Nueva serie*, s/f). Si los mismos datos se tomaran en deciles, el máximo ingreso sería 163,282.00 (valor promedio del decil más alto). En este caso $x_{MIDE} = 50{,}308.9991$, $IDRM = 0.6919$, $\tau = 2.25$ y $U\left(\underline{x}_{()}\right) = 0.3081$. Los últimos tres valores difieren mucho de los anteriores por ser calculados con valores agrupados en grupos muy grandes, aunque la diferencia singular es la del ingreso máximo.

### *El número de Palma (P) y el IDRM*

#### *El IDRM como función del número de Palma*

De la ecuación $(nt)_{\underline{x}} = (nt)_{\underline{x}_{()}}$, se tiene:

$$(nt)_{\underline{x}} = (nt)_{\underline{x}_{()}} = \sum_{i=1}^{n} p_{(i)} \left( \frac{x_{(n)} - x_{(i)}}{x_{(n)}} \right)$$

Puesto que el número de Palma utiliza los ingresos de los deciles, aquí $n = 10$, los $p_{(i)} = 0.1, i = 1, 2, \dots, 10$ y un número de Palma es $P = x_{(10)}/x_{(1)}$ cuando se toma el decil superior contra el decil inferior. No obstante, note que el sencillo desarrollo siguiente puede llevarse con cualquier número de Palma $P = x_{(n)}/x_{(1)}$ si en la *n*-ésima categoría se considera un porcentaje (ej. 10%) de los de mayor ingreso y en la primera categoría se considera un porcentaje (ej. 40%) de los de menor ingreso. De lo anterior:

$$(nt)_{\underline{x}} = (nt)_{\underline{x}_{()}} = \sum_{i=1}^{n} p_{(i)} \left(\frac{x_{(n)} - x_{(i)}}{x_{(n)}}\right) = \sum_{i=1}^{n} p_{(i)} \left(\frac{(x_{(n)} - x_{(i)})/x_{(1)}}{x_{(n)}/x_{(1)}}\right) = \sum_{i=1}^{n} p_{(i)} \left(\frac{P - x_{(i)}/x_{(1)}}{P}\right) =$$

$$\sum_{i=1}^{n} p_{(i)} - P^{-1} \sum_{i=1}^{n} p_{(i)}\, x_{(i)}/x_{(1)} = \sum_{i=1}^{n} p_{(i)} - P^{-1}\left(p_{(1)} + \sum_{i=2}^{n-1} p_{(i)}\, x_{(i)}/x_{(1)} + p_{(n)} P\right) =$$

$$\sum_{i=1}^{n} p_{(i)} - p_{(n)} - P^{-1}\left(p_{(1)} + \sum_{i=2}^{n-1} p_{(i)}\, x_{(i)}/x_{(1)}\right) = 1 - p_{(n)} - P^{-1}\left(p_{(1)} + \sum_{i=2}^{n-1} p_{(i)}\, x_{(i)}/x_{(1)}\right)$$

(3)

Note que la expresión entre paréntesis del lado derecho de la última ecuación no depende de *P*, ya que, es independiente de $x_{(n)}$, la denotaremos por *P**. Así, el *IDRM* está en función de *P* y *P**, el resto es una constante.

***Dado un valor de P, P* explica lo que deja de explicar P en el IDRM***

Considere un valor fijo de *P* en (3), es decir, $x_{(n)}$ y $x_{(1)}$ están fijos. Los ingresos $x_{(i)}, i = 2,3, \dots, n-1$, están libres y modifican al *IDRM*. Si $x_{(i)} = x_{(n)}, \forall i = 2,3, \dots, n-1$, entonces *P** toma su máximo valor o el *IDRM* su mínimo valor, al que denotaremos por $\underline{IDRM}$, y es:

$$\underline{IDRM} = 1 - p_{(n)} - P^{-1}\left(p_{(1)} + \frac{\sum_{i=2}^{n-1} p_{(i)}\, x_{(n)}}{x_{(1)}}\right) = 1 - p_{(n)} - P^{-1}\left(p_{(1)} + \sum_{i=2}^{n-1} p_{(i)}\, P\right)$$
$$= p_{(1)}(1 - P^{-1})$$

De manera análoga si $x_{(i)} = x_{(1)}, \forall\, i = 2,3, \dots, n-1$, entonces *P** toma su mínimo valor o el *IDRM* su máximo valor, al que denotaremos por $\overline{IDRM}$, y es:

$$\overline{IDRM} = 1 - p_{(n)} - P^{-1}\left(p_{(1)} + \frac{\sum_{i=2}^{n-1} p_{(i)}\, x_{(1)}}{x_{(1)}}\right) = 1 - p_{(n)} - P^{-1}\left(p_{(1)} + \sum_{i=2}^{n-1} p_{(i)}\right)$$
$$= (1 - p_{(n)})(1 - P^{-1})$$

Así, el valor del *IDRM*, $(nt)_{\underline{x}} = (nt)_{\underline{x}_{()}}$, yace entre estos dos umbrales, es decir:

$$\underline{IDRM} \leq (nt)_{\underline{x}_{()}} \leq \overline{IDRM}$$

Una medida de lo que deja de explicar *P* en la medida de desigualdad (en este caso el *IDRM*) y que denotaremos por *NoP* estaría dada por:

$$NoP = \frac{(nt)_{\underline{x}_{()}} - \underline{IDRM}}{\overline{IDRM} - \underline{IDRM}} = \frac{(nt)_{\underline{x}_{()}} - p_{(1)}(1 - P^{-1})}{(1 - p_{(1)} - p_{(n)})(1 - P^{-1})}$$

*El IDRM es sensible al P*

Sean $\underline{y}_{(\,)} = \{y_{(i)}\}_{i=1,2,\ldots,n}$ y $\underline{x}_{(\,)} = \{x_{(i)}\}_{i=1,2,\ldots,n}$ dos distribuciones de ingresos tales que $\frac{y_{(i)}}{y_{(1)}} = \frac{x_{(i)}}{x_{(1)}}\ \forall\ i = 2, 3, \ldots, n-1$; $P_y = \frac{y_{(n)}}{y_{(1)}} > \frac{x_{(n)}}{x_{(1)}} = P_x$; y las proporciones $p_{(i)}$ son las mismas para toda *i* = 1, 2, …, *n* en las dos distribuciones, entonces es claro que $(nt)_{\underline{y}_{(\,)}} > (nt)_{\underline{x}_{(\,)}}$, es decir, si las razones respecto a los mínimos de los ingresos ajenos a *P* y las proporciones no cambian, a un mayor *P* se tiene un mayor *IDRM*.

## 4. Resultados y discusión

Abordaremos la discusión a partir de los resultados obtenidos y algunos ejemplos. En primera instancia veremos una comparación del índice propuesto con las medidas tradicionales de desigualdad y posteriormente valoraremos el desempeño práctico del *IDRM*.

### 4.1. Propiedades deseables

La Tabla 2 muestra el desempeño comparativo de las medidas de desigualdad más frecuentes en cuanto a las propiedades que se han establecido como deseables. La tabla incluye al índice propuesto (*IDRM*) y la familia de razones de Palma, cuyo uso se ha extendido entre los estudios que describen desigualdad, por su interpretación accesible. Al considerar las nueve propiedades teóricas listadas en la sección 3.2, con valoraciones igualmente importantes de cumplimiento, se observa que el *IDRM* es tan robusto como el resto de los índices que se comparan (todos los índices cumplen ocho propiedades, excepto Dalton y Gini, Palma no es un índice de desigualdad). Cowell (2011) sugiere, si se desea considerar a todas y cada una de las distribuciones del ingreso posibles (no alguna familia de distribuciones con características particulares) que es razonable exigir se cumplan el principio de transferencias de Dalton-Pigou (fuerte) y la propiedad de descomponibilidad. Todos los índices que se comparan cumplen el principio de transferencias de Dalton-Pigou (fuerte), a excepción del *IDRM*, cumple la condición débil, lo cual puede ser valorado como una ventaja en la medición de la desigualdad y no una debilidad. Para ver lo anterior, considere una distribución de ingresos con 5 individuos A, B, C, D y E, sus ingresos respectivos son 0, 10, 25, 50 y 80 unidades, el promedio de ingresos es 33; el índice de Gini (más usado para analizar desigualdad) toma el valor de 0.485, el *IDRM* toma el valor de 0.588. Considere una transferencia de 4 unidades del individuo B al individuo A, la nueva distribución es: 4, 6, 25, 50, 80; Gini toma el valor de 0.475 e indica una disminución de la desigualdad, *IDRM* mantiene su valor. En efecto, hay una disminución de la desigualdad entre A y B, ahora están más cerca uno de otro en la escala social, sin embargo, observar que A está en una “mejor” condición, “más cercana” a los restantes cuatro individuos, ello ocurrió a costa de la transferencia obtenida de B, que ahora está en mayor desigualdad respecto a C, D y E. Por su parte la concentración de los tres individuos más ricos se mantuvo. En este escenario sería difícil admitir que la desigualdad ha disminuido (como lo señalarían todos los índices, a excepción del *IDRM*) y surge la duda de si la propiedad del principio de transferencias fuerte tiene la capacidad de capturar la idea general sobre desigualdad (Chateauneuf & Moyes, 2005). El *IDRM* cumple siempre la condición débil del principio de transferencias, señalando que la desigualdad se mantiene y cumple la condición fuerte sólo si la transferencia ocurre desde el individuo con el valor máximo, en cuyo caso habría una reducción de la desigualdad, no sólo entre el individuo que recibe la transferencia y el máximo, también entre el resto de los individuos y el máximo[4].

Observar también que si el individuo B, en la distribución original, se encuentra ligeramente por arriba de alguna línea hipotética de pobreza, la transferencia podría colocarlo en tal condición sin una mejora sustantiva en el individuo A. Retomando a Cowell (2011), sería ideal contar con una manera de ver la desigualdad que refleje tanto la profundidad de la pobreza de los "pobres" de la sociedad como el nivel de bienestar de los "ricos": no es fácil lograr esto simplemente mirando los ingresos acumulados o la riqueza que poseen dos o tres grupos. En este sentido, proponemos al *IDRM* como una herramienta robusta para atender este ideal.

**Tabla 2. ¿Qué cumple cada medida?**

| **índices de desigualdad & propiedades** | **Independencia de escala & tamaño de población (P2 & P3)** | **Principio de transferencias (P4 & P5)** | **Rango en intervalo $[0,1]$ (P6, P7 & P8)** | **Descomponible (P9)** |
|---|---|---|---|---|
| Atkinson (8) | Sí | Fuerte P4 | Sí | Sí |
| Dalton (6) | No | Fuerte P4 | No | Sí |
| Entropía Gen. (8) | Sí | Fuerte P5 | No | Sí |
| Gini (7) | Sí | Fuerte P4 | Sí | No |
| Herfindahl (8) | No | Fuerte P5 | Sí, pero $min > 0$ | Sí |
| MLD (8) | Sí | Fuerte P5 | No | Sí |
| Theil (8) | Sí | Fuerte P5 | No | Sí |
| IDRM (8*) | Sí | Débil P4* & Débil P5 | Sí | Sí |
| Palma ($\boldsymbol{T_x}/\boldsymbol{B_y}$) (3) | Sí | No** | No | No |

Notas: Todas las medidas de desigualdad cumplen el principio de anonimidad o simetría (P1). ***** el *IDRM* es invariante a transferencias de ingreso entre unidades, siempre que la unidad que transfiere no sea el máximo, así siempre cumple el principio de Dalton-Pigou de manera débil; si la transferencia se realiza desde el máximo, el principio se cumple de manera fuerte. ** Un cambio en la razón de Palma ocurriría si la transferencia ocurre del decil superior $\boldsymbol{T_x}$, al decil inferior $\boldsymbol{B_y}$.
Fuente: Elaboración propia, con base en Cowell (2011, pág. 72) y desarrollos propios.

[4] Notar que otra estrategia para disminuir la desigualdad consistiría en aumentar el ingreso de todos los individuos adicionando una constante $k > 0$, incluido el máximo.

### 4.2. Contraste del *IDRM* con otras medidas

Se presentan resultados prácticos para ilustrar las bondades del índice propuesto y su contraste con las medidas estándares de medición de la desigualdad. Se utilizó la base de datos sobre desigualdad de ingresos en el mundo (UNU-WIDER, 2022) que a escala global es el conjunto más completo de estadísticas sobre desigualdad de ingresos disponible, la versión utilizada (30 de junio de 2022) contiene información de 201 países y algunas cifras hasta el año 2021.

Se seleccionaron cifras de los ingresos netos per cápita en los hogares de cincuenta y ocho países de los cinco continentes, con la característica de tener series de datos con al menos tres mediciones cubriendo un periodo de al menos 5 años. La Tabla 1 del Anexo muestra los países seleccionados, el periodo disponible de información, el reporte del Índice de Gini, la razón de Palma (10) y el cálculo del *IDRM* en año inicial y final del periodo. También se reporta el porcentaje de cambio (anualizado en el periodo disponible) que ocurrió del año inicial (Año i) al último año disponible en la base de datos (Año f).

***Congruencia***

El primer contraste del *IDRM* se realizó considerando a los índices de Gini, Atkinson y Theil, medidas preferentes en los análisis de desigualdad y con importancia reciente al indicador denominado razón de Palma. Al ser estos índices e indicador diferentes medidas sobre la misma distribución del ingreso (en cada país), se espera que sus valores expresen entre sí algún patrón de correlación y que las discrepancias en el valor medido puedan atribuirse a las propiedades inherentes a cada índice. Lo anterior se verificó midiendo la asociación lineal (coeficiente de correlación de Pearson) entre los valores de cada índice, referentes al último año de información disponible. Los resultados se reportan en la matriz de correlaciones A (Tabla 3): tres índices de desigualdad (Gini, Atkinson y Theil) muestran una asociación lineal casi perfecta (correlaciones superiores a 0.982) mostrando *grosso modo* que cualquier elección de medida produciría en esencia los mismos resultados. Considerando a estos tres índices referentes, las mediciones de *IDRM* también muestran asociaciones fuertes como esperado, pero incorporan discrepancias con los panoramas de desigualdad referentes, las correlaciones están entre 0.925(Theil) y 0.969(Gini). Sobre el indicador de Palma, cuyo uso se ha extendido por proporcionar una idea intuitiva e interpretación directa de la desigualdad (Mejía & Chaparro, 2020), es la medida cuyos resultados describen la menor asociación con los cuatro índices y siendo la menor, con el *IDRM*; no olvidar que dicho indicador, sacrifica en su fin aspectos formales de la medición.

Un segundo contraste tiene que ver con el panorama que muestra cada índice sobre el cambio de la desigualdad en el tiempo (ver matriz de correlación B): similar a lo anterior, las mediciones de Gini, Atkinson y Theil señalan cambios temporales altamente correlacionados (disminuciones o aumentos de la desigualdad en los países), el indicador de Palma y el *IDRM* proporcionan resultados en los cambios de desigualdad menos correlacionados o distintos que los que describen los referentes usuales.

En síntesis, *IDRM* proporciona resultados congruentes (correlacionados) con los panoramas de desigualdad que describen los índices usuales, sin embargo, incorpora en su diseño aspectos como el Principio de Transferencias Dalton-Pigou (condición débil) que proporciona un efecto robusto ante transferencias no significativas (en términos de desigualdad) de ingresos, y ello puede conducir a que los cambios de la desigualdad en el tiempo presenten resultados distintos a los índices convencionales.

**Tabla 3. Matriz de correlación en variables seleccionadas**

| | A - Medidas de desigualdad año f | | | | |
|---|---|---|---|---|---|
| | Gini | Atkinson | Theil | IDRM | Palma |
| **Gini** | 1.000 | | | | |
| **Atkinson** | 0.992 | 1.000 | | | |
| **Theil** | 0.982 | 0.992 | 1.000 | | |
| **IDRM** | 0.969 | 0.935 | 0.925 | 1.000 | |
| **Palma10** | 0.832 | 0.883 | 0.896 | 0.732 | 1.000 |

| B – Mediciones de cambio % | | | | |
|---|---|---|---|---|
| Gini | Atkinson | Theil | IDRM | Palma |
| 1.000 | | | | |
| 0.978 | 1.000 | | | |
| 0.997 | 0.967 | 1.000 | | |
| 0.896 | 0.811 | 0.920 | 1.000 | |
| 0.836 | 0.897 | 0.828 | 0.623 | 1.000 |

Fuente: Elaboración propia.

***Sensibilidad y magnitud***

Una visualización para analizar los datos se muestra en la Gráfica 2, se contrastaron el *IDRM* y el índice de Gini vs el indicador de Palma 10 para el último año disponible en la base de datos. Se omitieron 2 observaciones (Cote d'Ivoire y South Africa) cuyos valores para el indicador de Palma son superiores a 60 veces (proporción de ingreso del decil superior vs decil inferior) y para el índice de Gini e *IDRM* superiores a 0.5.

Para analizar la información, recordar que tanto el índice de Gini como *IDRM* toman valores en el intervalo [0,1] y que un valor cercano 0 indica una condición cercana a la distribución igualitaria de los recursos y un valor cercano a 1 indica una condición de extrema desigualdad, cuando un individuo o un conjunto reducido de individuos poseen la gran masa de recursos. Estas ideas inducen a su vez la idea de que valores por debajo de 0.5 están más cercanos a cero y por tanto a una condición más cercana a la igualdad y distante de la extrema desigualdad. Así, la media del índice de Gini para los países analizados corresponde a 36.6%, la mediana 34.43%, resultando que unos pocos países (5) se encuentran por encima del 50%. Una conclusión sintética y general sería que el conjunto de países analizados tiene condiciones de desigualdad que están más cercanas al escenario positivo ideal y sólo cinco países están más cerca de condiciones no deseables (Brazil, India, Colombia, Cote d'Ivoire y South Africa). El indicador de Palma, con el propósito de dibujar una idea sensibilizadora de lo que significa la desigualdad, muestra que la proporción de ingreso acumulada por el decil superior respecto al inferior es en promedio 17.69 veces, con mínimo de 5.39 y máximo de 105.45 veces; en contraste a Gini, Palma pone a luz la crudeza de la desigualdad: aún en el país menos desigual (Slovakia, índice de Gini 23.24%) la distancia entre los ingresos del primer decil y ultimo decil es poco superior a 5 veces. Sin embargo, queda sin responder qué ocurre en el 80% de la población comprendida entre el primer y último decil: ¿cómo se modifica la desigualdad?

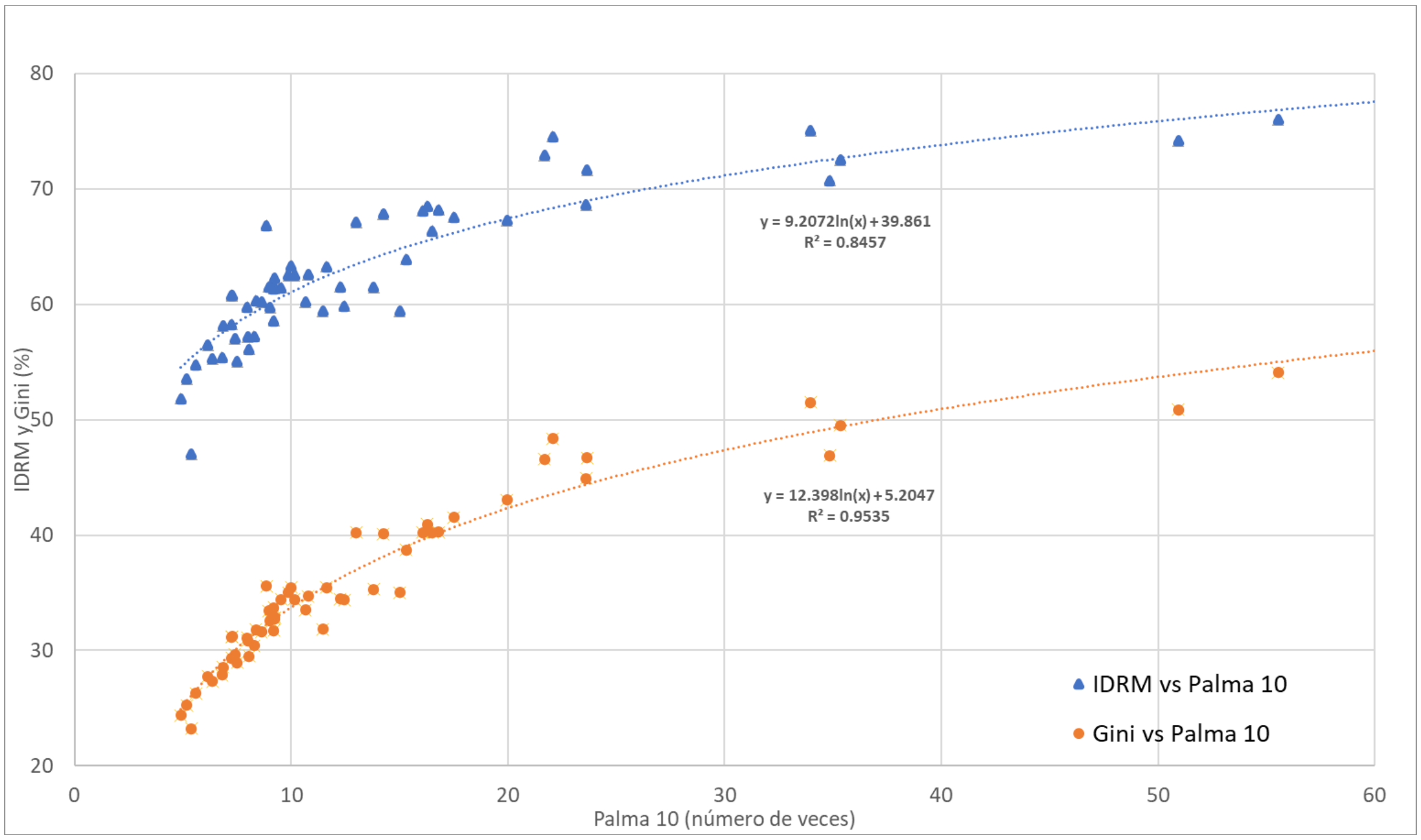

*Gráfica 2. Contrastes IDRM vs Palma 10 e Índice de Gini vs Palma 10.*

**Fuente. Elaboración propia.**

Por el lado de *IDRM* se observa que sólo un país (Slovakia, *IDRM* 47.00) se encuentra debajo del 50%, describe un escenario ligeramente más cercano a una condición de igualdad; el promedio de *IDRM* es 63.21, mediana 66.59 y máximo 82.40 (South Africa), muestra una visión congruente con Palma pero más robusta, al resaltar la agudeza de la desigualdad (el grueso de los países está más cerca del escenario no deseable) pero considerando el aporte a la desigualdad de lo que ocurre en los grupos intermedios de la población de los países, es decir, considerando toda la información. En resumen, se requiere que la razón entre la proporción que concentra el decil superior y la respectiva del decil inferior sea de al menos 30 veces para que Gini tome valores ligeramente superiores al 50%. De esta inspección es posible concluir en dos aspectos, la sensibilidad de *IDRM* para mostrar que la mayoría de los países están más cerca de la extrema desigualdad (valores superiores a 50%) y la magnitud de la desigualdad, ya que en todos los países donde Palma10 es superior a 30, *IDRM* toma valores superiores a 70%.

Observar que Gini y Palma guardan una relación más estrecha en su descripción de la desigualdad (pseudo R^2=0.95) que la relación que describe *IDRM* vs Palma (pseudo R^2=0.84), siendo la explicación de estas variaciones que Gini muestra poca variación en el 80% de la población ubicada entre los deciles extremos, como Palma, a diferencia del *IDRM*.

Sobre los cambios anualizados, tanto de incremento como de decremento en desigualdad (del año inicial al año final) que señala Palma tienen correspondencia con los cambios que señala el *IDRM*, con la anotación de que los cambios señalados por *IDRM* son de menor magnitud, tanto para indicar una distribución más igualitaria como para señalar la agudización de la desigualdad. Por ejemplo, para el caso de Peru, Palma señala un decremento en desigualdad (2004 a 2016) de -8.73 y en el caso de *IDRM* de apenas -0.60, es decir, un escenario donde la desigualdad prácticamente se ha mantenido constante. Otro ejemplo es Japan, donde el cambio anual señalado por Palma en el periodo 2009-2014 indica una disminución de la desigualdad en -4.35%, por su parte *IDRM* indica un incremento de la desigualdad de 0.15%. Es notorio que los estudios de desigualdad deberían realizarse con las herramientas robustas que sean congruentes con la realidad de las personas. Postulamos de estos análisis descriptivos que *IDRM* podría fungir como una herramienta sensible y consistente para el monitoreo de la desigualdad.

***Sesgo sobre datos agrupados***

De manera frecuente los índices o indicadores empleados para medir la desigualdad utilizan para su cálculo datos agrupados, siendo lo más común el uso de quintiles o deciles, esto tiene su justificación en que datos así agrupados son estadísticamente comparables entre países o bien los resultados agrupados tienen representatividad de la población objetivo, adicionalmente se tiene que las proporciones de ingreso en quintiles o deciles se manejan y comprenden con relativa facilidad.

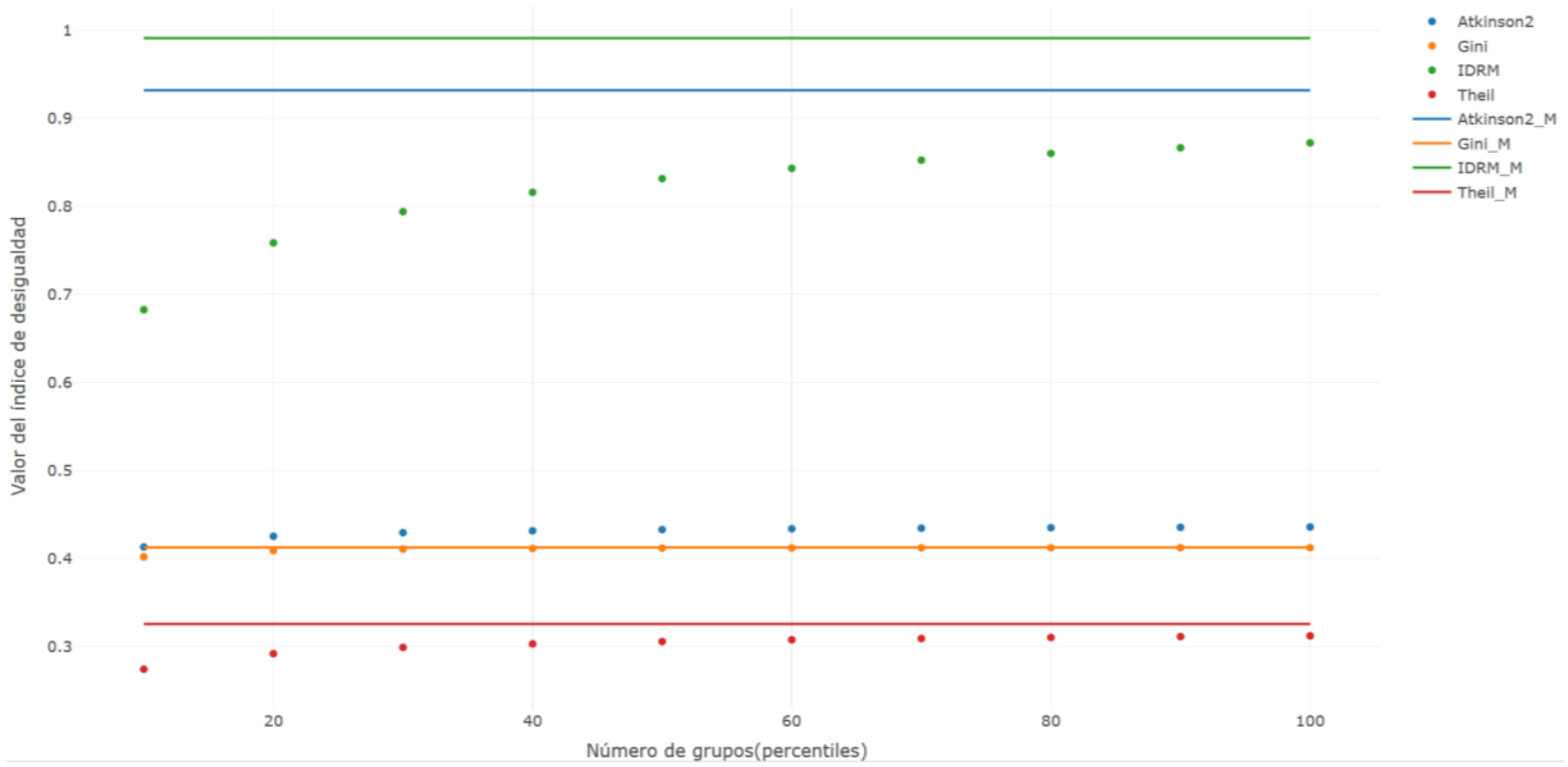

*Gráfica 3. Sesgo de cálculo de la desigualdad usando datos agrupados.*

**Fuente. Elaboración propia.**

Sobre el índice de Gini, se tiene evidencia de que su valor obtenido a partir de datos agrupados puede representar una medición sesgada de la desigualdad (Lerman & Yitzhaki, 1984, 1989). Entonces, como un aspecto más para valoración del *IDRM*, se estudió su desempeño considerando como referente a los microdatos de la ENIGH 2022 y comparando los valores obtenidos al construir grupos (10, 20, …, y 100); también se inspeccionaron los índices referentes. Los hallazgos se reportan en la Gráfica 3. Usando deciles vs microdatos las subestimaciones porcentuales de Gini, Theil, *IDRM* y Atkinson son 2.6, 15.7, 31.1 y 53.8%, respectivamente; estas diferencias se hacen más pequeñas conforme se suaviza la distribución de ingresos (incrementando el número de grupos). Gini y Theil (microdatos vs deciles 0.4126 vs 0.4018 y 0.3258 vs 0.2744, respectivamente) muestran los menores sesgos respecto a su cálculo con microdatos; ambos pares de valores describen que la distribución de ingresos es más cercana a la igualdad (valores menores a 0.5 para Gini y mucho menores a $log(10)$, para el caso de Theil[5]). Los valores de *IDRM* y Atkinson, microdatos vs deciles, son 0.9910 vs 0.6826 y 0.9318 vs 0.4131, respectivamente; observar que los valores de microdatos (superiores a 0.93 > 0.5) indican que la distribución está más cerca de la extrema desigualdad; la misma interpretación se obtiene de *IDRM* obtenido con datos agrupados (con cualquier número de grupos), no así con Atkinson cuyos valores en datos agrupados son menores a 0.5. Si bien *IDRM* presenta un sesgo de 15.7% con deciles respecto a microdatos, es el único índice que guarda consistencia entre las conclusiones que se derivan de ambos tipos de datos e inclina la balanza en dirección de la extrema desigualdad.

[5] El valor máximo del índice de Theil es $log(n)$.

***Estabilidad-consistencia***

Una posible crítica sobre el *IDRM* es su dependencia del valor máximo que se observa en la distribución, lo que aparentemente podría provocar inestabilidad o gran variación en las mediciones; sin embargo, la presencia de valores que se alejan en extremo del promedio ocurre frecuentemente en los estudios del ingreso, y por tal razón no podrían considerarse *outliers*. De hecho, existe evidencia de que personas u hogares en las partes altas de la distribución tienen ingresos subestimados, al menos para el caso mexicano estudiado a través de la ENIGH (Del Castillo, 2015; Esquivel, 2015; González, 2023; Reyes, Teruel y Morales, 2017).

Se valoró la estabilidad de *IDRM* y las medidas tradicionales a través de simulación *Bootstrap* (Wasserman, 2006), método no paramétrico (no asume alguna distribución subyacente para el ingreso) para estimar la varianza y distribución de indicadores. Con los microdatos de la ENIGH 2022 se instrumentó el método Bootstrap para generar 1000 muestras. Los resultados se despliegan en la Tabla 4.

**Tabla 4. Coeficientes de variación en el *IDRM* y en medidas tradicionales de desigualdad.**

| Característica | Microdatos | | | | Deciles | | | |
|---|---|---|---|---|---|---|---|---|
| | Gini | Theil | Atkinson* | *IDRM* | Gini | Theil | Atkinson* | *IDRM* |
| **Valor observado** | 0.41268 | 0.32574 | 0.93183 | 0.99110 | 0.40196 | 0.27440 | 0.41313 | 0.68263 |
| **Media b** | 0.41230 | 0.32490 | 0.92516 | 0.99049 | 0.40162 | 0.27385 | 0.41270 | 0.68221 |
| **Error estándar b** | 0.00201 | 0.00656 | 0.02900 | 0.00148 | 0.00182 | 0.00301 | 0.00241 | 0.00223 |
| **CV (%)** | 0.48733 | 2.01806 | 3.13492 | 0.14938 | 0.45357 | 1.09792 | 0.58507 | 0.32744 |

*En el índice de Atkinson se utilizó el parámetro de aversión $\varepsilon = 2$.
**Fuente: Elaboración propia.**

Del cálculo de los promedios Bootstrap, que por diseño del método se espera que aproximen a los valores observados, notamos que todas las estimaciones son cercanas a sus correspondientes observados (diferencias menores a 0.72%) y es posible concluir que el proceso *Bootstrap* generó muestras representativas de la población. En este escenario analizamos al coeficiente de variación (CV) a partir de las 1000 muestras. El CV expresa a la desviación estándar como un porcentaje de la media, es una medida relativa de precisión: conforme sus valores son más próximos a cero, la estimación es más precisa y viceversa (INEGI, 2017). Considerando la simulación con muestras *Bootstrap* y el CV obtenido a partir de microdatos se concluye que el *IDRM* es el índice más preciso y estable de los índices que se comparan (CV = 0.15 < 0.49 (Gini) < 2.01 (Theil) <3.13 (Atkinson)). Al analizar las precisiones obtenidas a partir de deciles, también se observa que el *IDRM* tiene la mejor precisión (más cercana a cero) y se verifica que el agrupamiento de los datos produce una menor precisión en el indicador (como se esperaría al disminuir un tamaño de muestra). Observar que las precisiones de Gini, Theil y Atkinson, al pasar de microdatos a deciles, aparentemente son mejores (contrario a *IDRM*, los CV disminuyen), lo cual es contradictorio y revela en realidad una subestimación del CV con deciles (o sobrevaloración de la precisión), esta situación advierte un riesgo de conclusiones erradas al usar deciles en los tres indicadores convencionales. Lo anterior refuerza la ventaja en estabilidad y precisión que tiene *IDRM* comparativamente con Gini, Theil y Atkinson.

### 4.3. Descomposición aditiva del *IDRM*

Una propiedad de importancia capital y valoración del *IDRM* es la descomposición aditiva en subgrupos de población, permite analizar la desigualdad como un todo y fraccionarla en partes que describen la desigualdad dentro de los grupos que conforman a la población y la desigualdad que existe entre dichos grupos. Cabe recordar que esta propiedad deseable para las medidas de desigualdad no se satisface por el índice de Gini, sí con el índice de Theil y Atkinson.

Para ilustrar la descomposición aditiva del *IDRM*, del conjunto de datos descrito en la sección 4.2 se seleccionaron registros de 43 países, cuyos datos estuvieron disponibles para 2018 y se analizaron como una población global, los resultados de la descomposición aditiva aparecen en la Tabla 5. En el conjunto de los 43 países considerados, el *IDRM* general toma el valor de 0.8993 señalando una condición general alejada de los ideales de igualdad y muy cercana a condiciones no deseadas de máxima desigualdad. Dicha desigualdad se compone por la desigualdad que priva al interior de cada continente (IW = 0.3837) y la desigualdad que existe por las disparidades entre los continentes (IB = 0.5156). En un segundo nivel de desagregación, se observa que América y Europa concentran las mayores proporciones de la población del conjunto y también de los ingresos totales; sin embargo, el continente americano contribuye con la mayor cantidad absoluta (IB = 0.3175) a la desigualdad general dentro de continentes (IW); en contraste, Europa a pesar de tener la mayor proporción poblacional y una proporción alta de ingresos, contribuye de manera marginal a la desigualdad (IB=0.0631), describiendo que los ingresos entre sus países están distribuidos más igualitariamente que en América.

En el tercer y último nivel de desagregación, la desigualdad de América se fracciona en la desigualdad entre sus países (IB = 0.1512) y la desigualdad al interior de los países (0.1662), las cifras indicarían que hay casi tanta desigualdad al interior de los países como entre ellos. Se observa que Estados Unidos concentra la proporción de 0.4493 del ingreso total del conjunto y 0.2370 de la población, contribuye con la mayor cantidad absoluta a la desigualdad entre países, resaltando que es un país de desigualdad extrema, posiblemente la más crítica del conjunto, considerando a su población y a su ingreso, simultáneamente.

Este ejemplo intenta clarificar y resaltar las bondades de la desagregación del *IDRM* en subgrupos de población para identificar y priorizar grupos donde las desigualdades sobresalen, tema que es de amplio interés en política pública.

**Tabla 5. Ejemplo. Descomposición del *IDRM* por subgrupos de población**

| Categoría | *IDRM* (IB+IW) | IB | IW | Proporción poblacional | Proporción de ingreso |
|---|---|---|---|---|---|
| **General** | | | | | |
| **43 Countries** | 0.8993 | 0.5156 | 0.3837 | 1.0000 | 1.0000 |
| **Continente** | | | | | |
| **Americas** | 0.3175 | 0.1512 | 0.1662 | 0.3988 | 0.5389 |
| **Europe** | 0.0631 | 0.0352 | 0.0278 | 0.4882 | 0.4102 |
| **Africa** | 0.0025 | 0.0005 | 0.0020 | 0.0851 | 0.0193 |
| **Oceania** | 0.0006 | 0.0000 | 0.0006 | 0.0180 | 0.0248 |
| **Asia** | 0.0001 | 0.0000 | 0.0001 | 0.0098 | 0.0067 |
| **País** | | | | | |
| **United States** | 0.1588 | | | 0.2370 | 0.4493 |
| **Mexico** | 0.0053 | | | 0.0914 | 0.0357 |
| **Canada** | 0.0013 | | | 0.0269 | 0.0355 |
| **Colombia** | 0.0009 | | | 0.0360 | 0.0144 |
| **Paraguay** | 0.0000 | | | 0.0050 | 0.0025 |
| **Uruguay** | 0.0000 | | | 0.0025 | 0.0015 |

**Fuente. Elaboración propia.**

### 4.4. Desigualdad de ingresos en México 2016-2022

Con el propósito de describir el panorama sobre desigualdad de ingresos en los hogares de México utilizando *IDRM*, se utilizaron datos de la Encuesta Nacional de Ingresos en los Hogares en México, colectados para el periodo 2016-2022 (INEGI, 2023), por ser estas mediciones directamente comparables entre sí[6]. La Tabla 6 muestra el ingreso corriente total promedio por hogar, por deciles e índices de desigualdad seleccionados

**Tabla 6. Ingreso corriente total promedio trimestral por hogar en deciles de hogares y mediciones sobre desigualdad.**

| | Año (pesos constantes 2022) | | | | Cambio |
|---|---|---|---|---|---|
| **Deciles / Índice** | **2016** | **2018** | **2020** | **2022** | **% (16-22)** |
| **Total promedio** | 63,565 | 60,916 | 57,370 | 63,695 | 0.20 |
| **I** | 11,141 | 11,183 | 11,333 | 13,411 | 20.38 |
| **II** | 19,382 | 19,755 | 19,229 | 22,421 | 15.68 |
| **III** | 25,811 | 26,288 | 25,400 | 29,201 | 13.13 |
| **IV** | 32,138 | 32,743 | 31,426 | 35,947 | 11.85 |
| **V** | 39,311 | 39,640 | 38,050 | 43,341 | 10.25 |
| **VI** | 47,537 | 47,777 | 45,737 | 51,924 | 9.23 |
| **VII** | 57,904 | 57,979 | 55,501 | 62,412 | 7.78 |
| **VIII** | 72,868 | 72,239 | 69,103 | 76,736 | 5.31 |
| **IX** | 98,333 | 96,445 | 91,726 | 100,866 | 2.58 |
| **X** | 231,226 | 205,106 | 186,198 | 200,696 | -13.20 |
| | | | | | |
| **Gini-deciles** | 0.449 | 0.426 | 0.415 | 0.402 | -10.48 |
| ***IDRM*-deciles** | 0.724 | 0.702 | 0.692 | 0.683 | -5.78 |
| **IDRM-micro** | 0.999 | 0.989 | 0.995 | 0.991 | -0.76 |
| **Palma** | 20.75 | 18.34 | 16.43 | 14.97 | -27.90 |

Fuente: Elaboración propia, a partir de los tabulados básicos de la ENIGH 2022 y cálculos propios.

La dinámica de la distribución del ingreso en México, y por tanto de la desigualdad, es altamente compleja de entender y más aún de intentar explicar: De 2016 a 2020 el ingreso promedio tuvo dos decrementos significativos -4.17% (2016-2018) y -5.82% (2018-2020). En el primer periodo el decremento se explica por un decremento del ingreso del último decil (-11.3%) y decrementos menores en los deciles IX y VIII, pues los primeros siete deciles tuvieron incrementos marginales, entre 0.13 y 1.92%. ¿Qué se puede decir sobre la desigualdad? El índice de Gini en deciles cambió de 0.449 a 0.426 (menor a 0.5, acercándose a una distribución más igual). *IDRM* disminuyó de 0.724 a 0.702 (una distribución que estaba más cercana la extrema desigualdad, mayor a 0.5, se atenuó sin cambiar la balanza). La desigualdad mejoró por la disminución de ingresos en el decil superior, y no como se hubiera esperado, por el progreso de los deciles inferiores. El *IDRM* con microdatos cambió de 0.999 a 0.989, señalando que una distribución extremadamente desigual se mitigó ligeramente.

[6] Ejercicios previos a 2016 de la ENIGH requieren tratamientos particulares para hacer comparables las series de datos, ver INEGI (2023).

En el segundo periodo (2018-2020) el decremento registrado en el promedio ocurrió por una disminución generalizada del ingreso en los deciles II a X, siendo este último el de mayor pérdida (-9,22%); estas disminuciones generalizadas podrían tener alta correlación con la severa parálisis del ciclo económico y la posterior contracción económica que impuso la pandemia por COVID-19 (González, 2023; Monroy, 2021). Se observa así una nueva distribución, con menores ingresos generalizados en los deciles II a X y prácticamente sin cambios en el decil I. Gini describe un cambio de 0.426 a 0.415 (-2.66%), un nuevo avance a una sociedad más igualitaria. *IDRM* por grupos también disminuye (-1.51%). La perspectiva es distinta al ver microdatos, esta nueva distribución, donde en general la población tiene menos ingresos, la desigualdad se incrementó.

En el periodo 2020-2022 el promedio total aumentó (11.03%), registrándose aumentos porcentuales generalizados en todos los deciles, desde 7.79% en el decil X hasta el máximo de 18.34% en el decil I. Gini e *IDRM* con deciles decrecen por tercera vez, disminuyendo 3.09 y 1.34% respectivamente. *IDRM* con microdatos también muestra una ligera mejora en la desigualdad (-0.42%).

En el periodo 2016-2022 el promedio de ingreso en los hogares se mantuvo prácticamente sin cambios (0.20% de incremento), los primeros seis deciles tuvieron incrementos superiores a 9% y el decil superior registró un decremento de 13%. El índice de Gini muestra un decremento de 10.48% con un valor final de 0.402, señala un avance significativo hacia la igualdad (0.402<0.5). *IDRM* decreció en el periodo 5.78% con un valor final de 0.683, indica avance menor respecto a Gini, pero aún en el ámbito contrario: de mayor cercanía a la desigualdad. *IDRM* con microdatos muestra que en esencia la desigualdad en el periodo se mantuvo constante (decremento de 0.76%), idea que parece la más congruente si se observan las curvas de Lorenz (Gráfica 4), que permiten comparar las distribuciones que se discuten: visualmente se observa un avance marginal en la línea azul (2022), respecto a la línea roja (2016); las distribuciones parecieran ser estáticas, lo cual tiene sentido en un periodo tan corto de análisis.

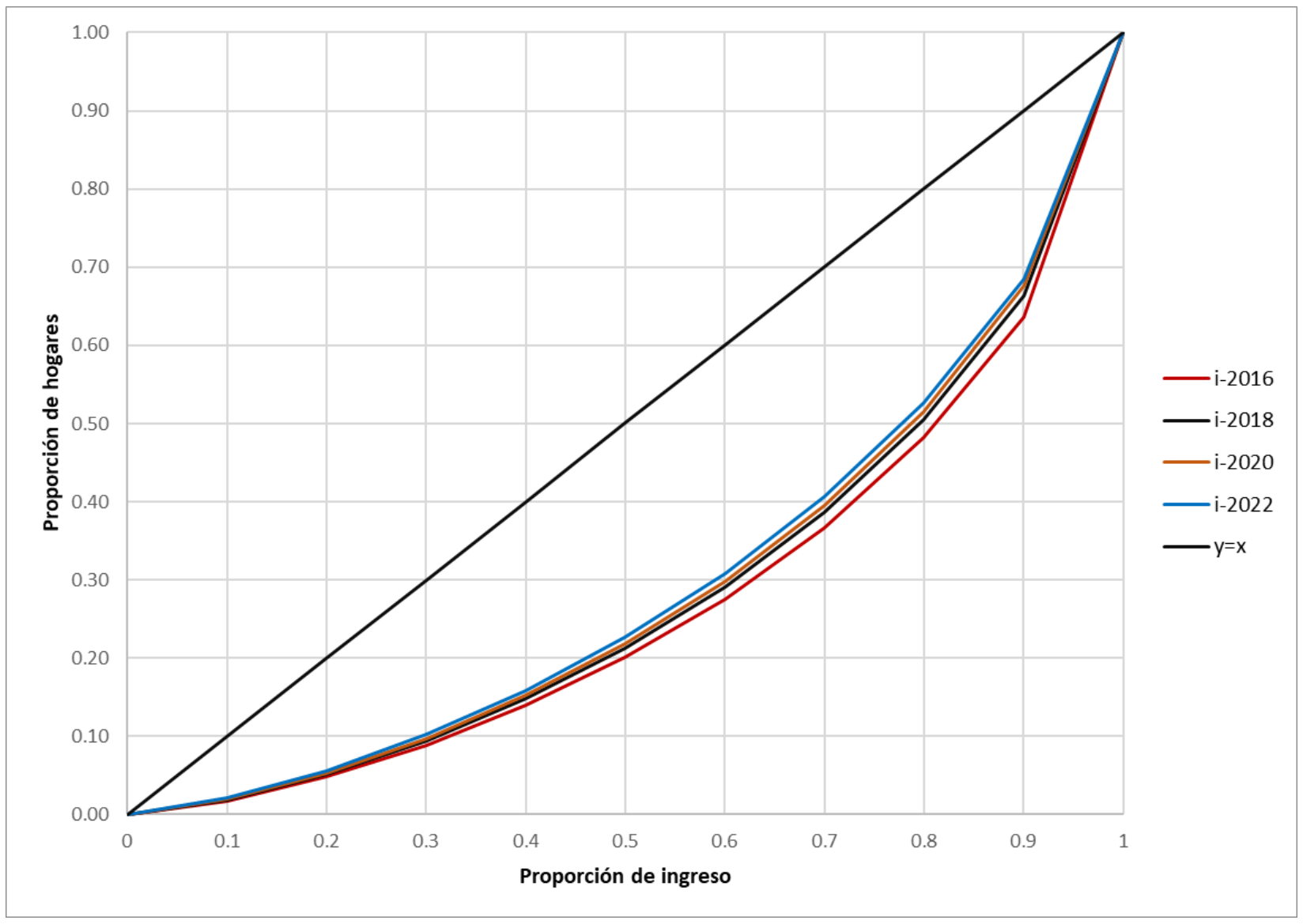

*Gráfica 4. Curvas de Lorenz para la distribución de ingreso en deciles de hogares 2016-2022.*

**Fuente. Elaboración propia.**

Varios estudios han mostrado la pertinencia de realizar ajustes a los datos obtenido de la ENIGH por subestimaciones de ingresos en deciles inferiores y superiores, ver por ejemplo a Del Castillo (2015), Esquivel (2015); González (2023) y Reyes, Teruel y Morales (2017). Del Castillo (2015) y González (2023) concluyen que, de corregirse las subestimaciones, nuevos valores se obtendrían para el índice de Gini, por ejemplo, pasar de valores en el rango 0.4-0.5 a valores en el rango 0.7-0.8, es decir, describiendo una distribución mucho más cercana a la desigualdad. *IDRM*, con deciles toma valores entre 0.68-0.72 y con microdatos toma valores superiores a 0.98, reflejando sin ningún ajuste en los datos que la distribución del ingreso con subestimaciones es una distribución altamente desigual.

## 5. Conclusiones

Presentamos propuesta de un índice para medir desigualdad. Realizamos el análisis del índice en dos perspectivas: a) respecto a las propiedades teóricas deseables que se establecen en la literatura científica para medidas de desigualdad (Cowell & Victoria-Feser, 1996). Observamos que es tan robusto como las medidas tradicionales de desigualdad y más robusto que el índice de Gini y Dalton: *IDRM* cumple 8 de las 9 propiedades, incluyendo la descomponibilidad en subgrupos de población (valorada con importancia capital en el análisis de desigualdad, ver Bourguignon (1979) y Cowell (2011)), propiedad que el índice de Gini, siendo el más utilizado, no satisface; b) respecto al desempeño empírico, usamos datos sobre distribución del ingreso de 58 países y microdatos de la distribución de ingreso en México (2016-2022) en un análisis comparativo de desempeño. Los índices de Gini, Theil y Atkinson se utilizaron como referentes y evaluamos la congruencia, sensibilidad, sesgo sobre datos agrupados y estabilidad-precisión de *IDRM*. Del análisis comparativo fue posible concluir que *IDRM* tiene un desempeño sobresaliente en la medición de la desigualdad respecto a los índices de referencia, significando por ello que captura y expresa de manera fácilmente interpretable las distintas formas que puede adoptar la distribución del ingreso, siendo además sensible a la magnitud de las desigualdades, característica de la que carecen los índices referentes.

Valoramos que el *IDRM* tiene importancia actual, pues varias investigaciones se han realizado con el propósito de visibilizar la magnitud de la desigualdad en consistencia con la realidad que se percibe. Los esfuerzos se han centrado en corregir estimaciones de la distribución del ingreso, por ejemplo, Del Castillo (2015), Esquivel (2015), González (2023), y Reyes, Teruel y Morales (2017), y no tanto en sensibilizar a las medidas de desigualdad que se usan. Postulamos a *IDRM* como una herramienta sensible a la profundidad de las desigualdades y con ello aproxima una idea del bienestar social, en el sentido que expresa Cowell (2011): capturando lo agudo de la pobreza y el nivel de bienestar de los ricos.

De los ejemplos expuestos, concluimos que *IDRM* tiene los elementos para postularse como una herramienta de política pública para el análisis de la desigualdad (tanto a escala macroeconómica, en comparativos internacionales, como en escalas locales, al interior de los países, esencialmente por la propiedad de descomponibilidad) y también para la investigación social, en temas donde frecuentemente se imbrican varios problemas, por ejemplo, desigualdad vs crecimiento económico, pobreza, desarrollo humano, acceso a derechos sociales, desigualdad ambiental, etcétera.

Algunas propiedades adicionales se podrían estudiar, por ejemplo, su descomposición en analogía con el análisis de varianza en estadística, de ser factible tal descomposición, permitiría analizar el efecto de los factores sobre una respuesta como expresión de la desigualdad que aquellos le provocan a ésta.

## 6. Bibliografía

## Anexo 1.

## Tabla 1. Índice de Gini, Razón de Palma e Índice de Desigualdad Relativo al Máximo, según país y años seleccionados.

| Continente | País | Periodo | | Gini | | | Theil | | | Atkinson (1) | | | IDRM | | | Palma10 | | |
|---|---|---|---|---|---|---|---|---|---|---|---|---|---|---|---|---|---|---|
| | | Año 1 | Año 2 | Año 1 | Año 2 | %D | Año 1 | Año 2 | %D | Año 1 | Año 2 | %D | Año 1 | Año 2 | %D | Año 1 | Año 2 | %D |
| **Africa** | Cote d'Ivoire | 2002 | 2015 | 58.79 | 60.67 | 0.24 | 59.65 | 63.78 | 0.52 | 48.02 | 51.83 | 0.59 | 78.71 | 79.34 | 0.06 | 92.12 | 138.31 | 3.18 |
| **Africa** | Egypt | 2000 | 2018 | 36.39 | 35.58 | -0.12 | 22.56 | 21.17 | -0.35 | 18.29 | 17.55 | -0.23 | 67.99 | 66.83 | -0.10 | 8.95 | 8.87 | -0.05 |
| **Africa** | Mali | 2012 | 2020 | 48.04 | 40.21 | -2.20 | 38.26 | 26.41 | -4.53 | 32.44 | 23.53 | -3.94 | 73.19 | 68.09 | -0.90 | 27.83 | 16.07 | -6.64 |
| **Africa** | South Africa | 2008 | 2017 | 69.72 | 66.99 | -0.44 | 87.43 | 79.45 | -1.06 | 63.82 | 58.67 | -0.93 | 82.12 | 81.40 | -0.10 | 159.77 | 105.45 | -4.51 |
| **Americas** | Brazil | 2006 | 2016 | 53.51 | 50.86 | -0.51 | 48.30 | 43.16 | -1.12 | 39.58 | 38.32 | -0.32 | 76.03 | 74.17 | -0.25 | 40.90 | 50.93 | 2.22 |
| **Americas** | Canada | 1972 | 2018 | 37.37 | 32.58 | -0.30 | 22.23 | 16.65 | -0.63 | 21.96 | 15.86 | -0.71 | 63.06 | 59.74 | -0.12 | 16.21 | 9.03 | -1.26 |
| **Americas** | Chile | 1990 | 2017 | 52.81 | 48.35 | -0.33 | 47.67 | 39.86 | -0.66 | 37.89 | 31.56 | -0.68 | 76.36 | 74.54 | -0.09 | 35.85 | 22.07 | -1.78 |
| **Americas** | Colombia | 2001 | 2020 | 54.27 | 54.09 | -0.02 | 49.83 | 49.10 | -0.08 | 41.72 | 41.56 | -0.02 | 76.33 | 76.00 | -0.02 | 57.08 | 55.55 | -0.14 |
| **Americas** | Mexico | 1984 | 2018 | 47.28 | 46.57 | -0.04 | 36.66 | 36.14 | -0.04 | 31.24 | 29.89 | -0.13 | 72.18 | 72.89 | 0.03 | 22.75 | 21.70 | -0.14 |
| **Americas** | Panama | 2007 | 2016 | 51.82 | 49.51 | -0.51 | 44.24 | 40.08 | -1.09 | 38.74 | 35.84 | -0.86 | 73.96 | 72.53 | -0.22 | 40.00 | 35.34 | -1.37 |
| **Americas** | Paraguay | 1998 | 2020 | 56.49 | 46.68 | -0.86 | 53.53 | 35.57 | -1.84 | 46.48 | 30.98 | -1.83 | 76.51 | 71.62 | -0.30 | 78.85 | 23.65 | -5.33 |
| **Americas** | Peru | 2004 | 2016 | 56.15 | 46.90 | -1.49 | 52.73 | 35.68 | -3.20 | 48.11 | 33.55 | -2.96 | 76.05 | 70.72 | -0.60 | 104.38 | 34.86 | -8.73 |
| **Americas** | United States | 1975 | 2020 | 35.66 | 41.54 | 0.34 | 20.14 | 27.60 | 0.70 | 19.85 | 25.28 | 0.54 | 61.76 | 67.55 | 0.20 | 13.55 | 17.51 | 0.57 |
| **Americas** | Uruguay | 2004 | 2019 | 47.24 | 40.22 | -1.07 | 36.45 | 25.92 | -2.25 | 31.22 | 22.99 | -2.02 | 71.96 | 67.12 | -0.46 | 21.88 | 13.00 | -3.41 |
| **Asia** | China | 2003 | 2014 | 44.68 | 43.04 | -0.34 | 31.86 | 29.39 | -0.73 | 28.55 | 27.71 | -0.27 | 68.92 | 67.26 | -0.22 | 18.07 | 19.96 | 0.91 |
| **Asia** | Cyprus | 2005 | 2020 | 30.09 | 31.22 | 0.25 | 14.28 | 15.61 | 0.60 | 13.03 | 14.02 | 0.49 | 58.86 | 60.77 | 0.21 | 6.66 | 7.28 | 0.60 |
| **Asia** | Georgia | 2009 | 2019 | 47.54 | 40.22 | -1.66 | 36.65 | 25.84 | -3.44 | 33.23 | 24.06 | -3.18 | 71.39 | 66.34 | -0.73 | 30.93 | 16.51 | -6.09 |
| **Asia** | India | 2005 | 2012 | 50.59 | 51.51 | 0.26 | 42.54 | 44.45 | 0.63 | 35.39 | 36.60 | 0.48 | 74.35 | 75.06 | 0.14 | 30.94 | 33.97 | 1.34 |
| **Asia** | Iraq | 2007 | 2013 | 41.20 | 40.92 | -0.11 | 28.05 | 27.13 | -0.55 | 23.88 | 24.05 | 0.12 | 69.73 | 68.49 | -0.30 | 15.51 | 16.28 | 0.81 |
| **Asia** | Israel | 1980 | 2018 | 36.34 | 38.70 | 0.17 | 20.71 | 23.61 | 0.35 | 19.14 | 23.00 | 0.48 | 62.94 | 63.90 | 0.04 | 10.14 | 15.30 | 1.09 |
| **Asia** | Japan | 2009 | 2014 | 34.84 | 32.88 | -1.15 | 19.24 | 17.51 | -1.87 | 18.51 | 15.93 | -2.96 | 61.80 | 62.26 | 0.15 | 11.53 | 9.23 | -4.35 |
| **Asia** | Jordan | 2003 | 2014 | 41.76 | 40.10 | -0.37 | 28.62 | 26.05 | -0.85 | 24.43 | 22.93 | -0.58 | 69.65 | 67.82 | -0.24 | 14.98 | 14.26 | -0.45 |
| **Asia** | Korea, Republic of | 2006 | 2016 | 31.97 | 31.72 | -0.08 | 16.08 | 15.82 | -0.16 | 15.80 | 15.42 | -0.24 | 58.52 | 58.55 | 0.01 | 9.53 | 9.21 | -0.34 |
| **Asia** | Vietnam | 2006 | 2014 | 37.78 | 35.45 | -0.79 | 23.35 | 19.94 | -1.95 | 19.88 | 18.21 | -1.09 | 67.06 | 63.35 | -0.71 | 10.29 | 9.99 | -0.37 |
| **Asia** | West Bank and Gaza | 2011 | 2017 | 44.82 | 44.90 | 0.03 | 32.21 | 32.17 | -0.02 | 29.07 | 30.19 | 0.63 | 69.47 | 68.64 | -0.20 | 20.86 | 23.61 | 2.09 |
| **Europe** | Austria | 1987 | 2020 | 23.05 | 30.40 | 0.84 | 8.14 | 14.50 | 1.76 | 7.94 | 14.14 | 1.76 | 48.22 | 57.21 | 0.52 | 4.40 | 8.29 | 1.94 |
| **Europe** | Belgium | 1985 | 2020 | 25.22 | 27.32 | 0.23 | 9.84 | 11.74 | 0.51 | 9.50 | 11.22 | 0.48 | 51.62 | 55.26 | 0.19 | 5.15 | 6.37 | 0.61 |
| **Europe** | Bulgaria | 2007 | 2020 | 36.10 | 40.27 | 0.84 | 20.99 | 26.53 | 1.82 | 21.48 | 23.67 | 0.75 | 62.48 | 68.17 | 0.67 | 17.65 | 16.80 | -0.38 |
| **Europe** | Croatia | 2010 | 2020 | 32.56 | 28.93 | -1.18 | 16.73 | 13.07 | -2.44 | 16.39 | 12.97 | -2.31 | 58.96 | 55.04 | -0.69 | 9.74 | 7.51 | -2.57 |
| **Europe** | Czechia | 1993 | 2020 | 20.71 | 25.27 | 0.74 | 6.84 | 10.09 | 1.45 | 6.38 | 9.45 | 1.46 | 48.98 | 53.51 | 0.33 | 3.84 | 5.18 | 1.11 |
| **Europe** | Denmark | 1987 | 2020 | 26.48 | 28.53 | 0.23 | 11.02 | 13.03 | 0.51 | 10.68 | 12.03 | 0.36 | 53.90 | 58.13 | 0.23 | 6.23 | 6.86 | 0.29 |
| **Europe** | Estonia | 2000 | 2020 | 36.64 | 30.87 | -0.85 | 21.70 | 14.89 | -1.87 | 19.65 | 14.35 | -1.56 | 65.10 | 57.17 | -0.65 | 12.24 | 8.00 | -2.11 |
| **Europe** | Finland | 1987 | 2020 | 22.19 | 27.74 | 0.68 | 7.60 | 12.16 | 1.43 | 7.36 | 11.32 | 1.32 | 47.97 | 56.48 | 0.50 | 4.21 | 6.16 | 1.16 |
| **Europe** | France | 1970 | 2020 | 37.23 | 31.60 | -0.33 | 22.11 | 15.89 | -0.66 | 19.61 | 14.91 | -0.55 | 65.10 | 60.20 | -0.16 | 10.49 | 8.64 | -0.39 |

| Continente | País | Periodo | | Gini | | | Theil | | | Atkinson (1) | | | IDRM | | | Palma10 | | |
|---|---|---|---|---|---|---|---|---|---|---|---|---|---|---|---|---|---|---|
| | | Año 1 | Año 2 | Año 1 | Año 2 | %D | Año 1 | Año 2 | %D | Año 1 | Año 2 | %D | Año 1 | Año 2 | %D | Año 1 | Año 2 | %D |
| **Europe** | Germany | 1973 | 2019 | 29.95 | 31.79 | 0.13 | 14.11 | 16.01 | 0.28 | 13.09 | 14.87 | 0.28 | 58.11 | 60.31 | 0.08 | 6.86 | 8.37 | 0.43 |
| **Europe** | Greece | 1996 | 2020 | 37.19 | 33.51 | -0.43 | 21.95 | 17.64 | -0.91 | 21.14 | 17.18 | -0.86 | 63.54 | 60.21 | -0.22 | 14.29 | 10.65 | -1.22 |
| **Europe** | Hungary | 1992 | 2020 | 29.62 | 29.66 | 0.00 | 14.20 | 13.66 | -0.14 | 13.84 | 13.14 | -0.19 | 58.17 | 57.00 | -0.07 | 8.99 | 7.41 | -0.69 |
| **Europe** | Iceland | 2004 | 2018 | 26.89 | 26.27 | -0.17 | 11.52 | 10.89 | -0.41 | 10.76 | 10.22 | -0.37 | 55.87 | 54.77 | -0.14 | 6.04 | 5.61 | -0.53 |
| **Europe** | Ireland | 1988 | 2020 | 35.93 | 31.15 | -0.45 | 20.51 | 15.52 | -0.87 | 18.92 | 13.90 | -0.96 | 63.53 | 60.80 | -0.14 | 11.01 | 7.25 | -1.30 |
| **Europe** | Italy | 1986 | 2019 | 32.52 | 35.30 | 0.25 | 16.52 | 19.76 | 0.54 | 15.68 | 19.68 | 0.69 | 59.59 | 61.46 | 0.09 | 8.65 | 13.80 | 1.42 |
| **Europe** | Latvia | 2005 | 2020 | 36.61 | 34.69 | -0.36 | 21.78 | 19.19 | -0.84 | 20.16 | 17.85 | -0.81 | 65.00 | 62.59 | -0.25 | 13.87 | 10.78 | -1.67 |
| **Europe** | Lithuania | 2005 | 2019 | 37.09 | 35.44 | -0.32 | 22.13 | 20.10 | -0.68 | 21.03 | 18.73 | -0.82 | 64.20 | 63.26 | -0.11 | 14.32 | 11.63 | -1.48 |
| **Europe** | Luxembourg | 1985 | 2020 | 26.67 | 34.44 | 0.73 | 11.05 | 18.64 | 1.51 | 10.45 | 17.45 | 1.48 | 53.68 | 61.42 | 0.39 | 5.36 | 9.53 | 1.66 |
| **Europe** | Malta | 2007 | 2020 | 28.18 | 31.08 | 0.76 | 12.38 | 15.38 | 1.68 | 12.05 | 14.27 | 1.31 | 54.73 | 59.78 | 0.68 | 6.65 | 7.97 | 1.40 |
| **Europe** | Netherlands | 1983 | 2020 | 28.41 | 29.35 | 0.09 | 12.80 | 13.67 | 0.18 | 12.35 | 12.79 | 0.09 | 55.93 | 58.25 | 0.11 | 7.32 | 7.26 | -0.02 |
| **Europe** | Norway | 1979 | 2020 | 27.01 | 27.86 | 0.08 | 11.42 | 12.21 | 0.16 | 10.81 | 11.79 | 0.21 | 54.38 | 55.38 | 0.04 | 5.81 | 6.81 | 0.39 |
| **Europe** | Poland | 1986 | 2020 | 27.99 | 31.84 | 0.38 | 12.17 | 16.30 | 0.86 | 11.89 | 16.33 | 0.94 | 54.25 | 59.40 | 0.27 | 6.53 | 11.46 | 1.67 |
| **Europe** | Portugal | 1995 | 2020 | 38.90 | 32.76 | -0.68 | 24.39 | 17.24 | -1.38 | 23.30 | 15.92 | -1.51 | 65.60 | 61.53 | -0.26 | 17.94 | 9.22 | -2.63 |
| **Europe** | Romania | 2007 | 2020 | 39.80 | 35.04 | -0.98 | 25.31 | 19.52 | -1.98 | 24.26 | 20.51 | -1.28 | 65.68 | 59.41 | -0.77 | 17.55 | 15.02 | -1.19 |
| **Europe** | Russia | 2000 | 2020 | 42.68 | 35.04 | -0.98 | 29.74 | 19.41 | -2.11 | 27.05 | 17.91 | -2.04 | 69.43 | 62.52 | -0.52 | 22.87 | 9.88 | -4.11 |
| **Europe** | Serbia | 2006 | 2020 | 37.14 | 34.48 | -0.53 | 21.92 | 18.90 | -1.05 | 22.59 | 18.46 | -1.43 | 61.93 | 61.51 | -0.05 | 17.87 | 12.25 | -2.66 |
| **Europe** | Slovakia | 1993 | 2020 | 20.19 | 23.24 | 0.52 | 6.34 | 8.50 | 1.10 | 6.10 | 8.81 | 1.37 | 46.31 | 47.00 | 0.05 | 3.73 | 5.39 | 1.38 |
| **Europe** | Slovenia | 1997 | 2020 | 24.12 | 24.38 | 0.05 | 9.11 | 9.28 | 0.08 | 8.86 | 8.83 | -0.01 | 50.67 | 51.81 | 0.10 | 5.08 | 4.92 | -0.14 |
| **Europe** | Spain | 1981 | 2020 | 34.49 | 34.37 | -0.01 | 18.80 | 18.62 | -0.02 | 17.60 | 18.76 | 0.16 | 61.99 | 59.86 | -0.09 | 10.24 | 12.46 | 0.50 |
| **Europe** | Sweden | 1975 | 2020 | 24.36 | 29.49 | 0.43 | 9.10 | 13.62 | 0.90 | 8.95 | 13.44 | 0.91 | 48.71 | 56.10 | 0.31 | 4.86 | 8.05 | 1.13 |
| **Europe** | Switzerland | 1982 | 2019 | 36.08 | 33.42 | -0.21 | 20.84 | 17.63 | -0.45 | 18.29 | 16.27 | -0.32 | 65.04 | 61.53 | -0.15 | 9.87 | 8.96 | -0.26 |
| **Europe** | United Kingdom | 1969 | 2019 | 33.68 | 33.73 | 0.00 | 17.96 | 17.91 | 0.00 | 16.17 | 16.68 | 0.06 | 61.72 | 61.35 | -0.01 | 8.09 | 9.21 | 0.26 |
| **Oceania** | Australia | 1982 | 2018 | 31.32 | 34.41 | 0.26 | 15.30 | 18.79 | 0.57 | 14.88 | 17.34 | 0.43 | 57.17 | 62.52 | 0.25 | 8.40 | 10.14 | 0.53 |

Fuente: Elaboración propia, con base en datos de UNU-WIDER (2022) y cálculos propios.